\documentclass[pdflatex,iicol,sn-mathphys-num]{sn-jnl}

\usepackage{graphicx}%
\usepackage{multirow}%
\usepackage{amsmath,amssymb,amsfonts}%
\usepackage{amsthm}%
\usepackage{mathrsfs}%
\usepackage[title]{appendix}%
\usepackage{xcolor}%
\usepackage{textcomp}%
\usepackage{manyfoot}%
\usepackage{booktabs}%
\usepackage{algorithm}%
\usepackage{algorithmicx}%
\usepackage{algpseudocode}%
\usepackage{listings}%

\usepackage{textcomp}

\begin{document}

\title[Article Title]{A two-axes shear cell for rheo-optics}

\author[1]{\fnm{Marraffa} \sur{Chiara}}
\author*[1]{\fnm{Aime} \sur{Stefano}}\email{stefano.aime@espci.fr}
\affil*[1]{\orgdiv{Molecular, Macromolecular Chemistry, and Materials}, \orgname{ESPCI Paris}, \orgaddress{\street{10 rue Vauquelin}, \city{Paris}, \postcode{75005}, \country{France}}}

\abstract{We develop and test a rheo-optical platform based on a two-axes, parallel plates shear cell coupled to an optical microscope and a photon correlation imaging setup, for simultaneous investigation of the rheological response and the microscopic structure and dynamics of soft materials under shear. Each plate of the shear cell is driven by an air bearing linear stage, actuated by a voice coil motor. A servo control loop reading the plate displacement through a contactless linear encoder enables both strain-controlled and stress-controlled rheology. Simultaneous actuation of both linear stages enables both parallel and orthogonal superposition rheology. We validate the performance of our device in both oscillatory and transient rheological tests on a microgel soft glass, and we demonstrate its potential through orthogonal superposition rheology experiments. During steady-state flow, we reconstruct the strain field across the gap by tracking the motion of tracer particles, to check for slip or shear banding instabilities. At the same time, we measure the microscopic dynamics, both affine and non-affine, resolving them in space and time using photon correlation imaging.}


\maketitle

\section{Introduction}\label{sec:intro}

Complex fluids—including polymer solutions, surfactant phases, foams, emulsions, and colloidal suspensions—are widespread in both everyday life and industrial processes. 
Their rich rheological behavior plays a crucial role during manufacturing and in their end-use performance \cite{wagnerTheoryApplicationsColloidal2021}. 
Rheology provides key guidance in formulation and processing of such products by reducing the system complexity to a minimal set of mechanical observables: stress and strain, often reduced to their one-dimensional shear components \cite{larsonStructureRheologyComplex1999}. 
Optimizing material formulation and processing is usually a complex task, as the constitutive relationship relating stress and strain is the result of many interdependent physico-chemical processes. In addition, most real-life applications are typically characterized by more complex mechanical loading than that of standard rheology. 
This calls for advanced rheological tests and analysis to more fully capture their rich mechanical behavior and to help disentangle the different contributions to the storage and dissipation of elastic energy~\cite{donleyElucidatingOvershootSoft2020,chenMicroscopicDynamicsStress2020,vinuthaMemoryShearFlow2024,ederaDeformationProfilesMicroscopic2021,sethMicromechanicalModelPredict2011a,mohanMicroscopicOriginInternal2013a,kongLocalizedElasticityGoverns2022}. 
Advanced analysis of the rheology signal reveals the tensorial nature of the stress: commercial rheometers provide limited access to it, through the measurement of axial forces and normal stress differences \cite{pommellaRoleNormalStress2020,costanzoMeasuringAssessingFirst2018,gauthierNewPressureSensor2021}, and can be usefully combined with extensional rheology \cite{costanzoShearExtensionalRheology2016, roddCapillaryBreakupRheometry2005}, or extended through custom modifications of shear rheometers to achieve shear along two orthogonal axes~\cite{simmonsServocontrolledRheometerMeasurement1966,zeegersSensitiveDynamicViscometer1995,vermantOrthogonalSuperpositionMeasurements1997,vermantOrthogonalParallelSuperposition1998}, 
a task which motivated also the development of dedicated custom-made instruments~\cite{linMultiaxisConfocalRheoscope2014,blancRheologyDenseSuspensions2023}. 

While these results demonstrate that rheology experiments are rich and informative, and call for advanced protocols and analysis techniques, they also highlight that the rheological signal is the result of several microscopic processes, whose complexity goes far beyond what can be inferred from macroscopic stress and strains measurements alone. 
Indeed, one major challenge in rheology is that of \textit{tracing the observed macroscopic behavior back to its microscopic origin}. 
This is also the focus of extensive fundamental research across a range of fields, such as the physics of polymers, foam dynamics, biological fluids and tissues, and glass transition \cite{rubinsteinPolymerPhysics2003,JanmeyRheology2007,chenRheologySoftMaterials2010,binderGlassyMaterialsDisordered2011,cantatFoamsStructureDynamics2013}.
These research fields have motivated the development of techniques combining rheology with other techniques probing local material properties at various relevant length scales, as detailed in multiple reviews and textbooks on so-called rheo-optical, or more generally rheo-physical, combined techniques \cite{fullerOpticalRheometryComplex1995,sondergaardRheoPhysicsMultiphasePolymer1995,wagnerRheooptics1998,callaghanRheoNMRNuclearMagnetic1999,vermantFlowinducedStructureColloidal2005,mannevilleRecentExperimentalProbes2008,fullerComplexFluidFluidInterfaces2012,peuvrel-disdierApportTechniquesCouplees2013,koponenAnalysisIndustryRelatedFlows2020,isaQuantitativeImagingConcentrated2010}. 
Hereafter, we restrict the discussion to shear rheology, discussing benefits and challenges of developing rheo-optical techniques based on home-made shear cells, instead of adapting commercial rheometers.

Transparent tools for rheometers make it relatively easy to couple rheology to turbidity measurements~\cite{rangel-nafaileAnalysisStressinducedPhase1984,yanaseStructureDynamicsPolymer1991}, birefringence~\cite{asadaRheoopticalStudiesRacemic1980}, optical macro imaging \cite{mckinleyObservationsElasticInstability1991,sunAnomalousCrystallineOrdering2023}, and microscopy \cite{kingNewAccessoriesWeissenberg1975,arigoEffectsViscoelasticityTransient1997,meekerSlipFlowPastes2004,dimitriouRheoPIVShearbandingWormlike2012,r.sethHowSoftParticle2012,hawDirectObservationOscillatoryshearinduced1998,martinezCombinedRheometryImaging2020}. The need to resolve three-dimensional features and dynamics has particularly promoted the use of laser-sheet microscopy \cite{barentinFlowSegregationSheared2004,huKineticsMechanismShear2005,lerougeInterfaceDynamicsShearbanding2008}, 
eventually coupled to fluorescence \cite{rashediEngineeredTransparentEmulsion2020} for 3D reconstruction of particle dynamics \cite{goudoulasNonlinearitiesShearBanding2017}, a task also accomplished by coupling rheometers to confocal microscopes \cite{besselingQuantitativeImagingColloidal2009,schmollerCyclicHardeningBundled2010,basuNonaffineDisplacementsFlexible2011,sentjabrskajaCreepFlowGlasses2015,himanagamanasaExperimentalSignaturesNonequilibrium2014,koumakisTuningColloidalGels2015}. 
More advanced imaging techniques such as Optical Coherence Tomography~\cite{vanleeuwenHighflowvelocityShearrateImaging1999}, ultrasound velocimetry~\cite{mannevilleHighfrequencyUltrasonicSpeckle2004,gibaudInfluenceBoundaryConditions2008} and X-ray radiography \cite{deboeufImagingNonBrownianParticle2018} have also been coupled to ad-hoc modified versions of commercial rheometers. 
However, coupling demanding imaging techniques to commercial rheometers is usually challenging, and requires the development of custom-made imaging instruments. 
Conversely, rheometers are sometimes replaced by home-made shear cells deforming the sample on commercial imaging instruments such as bright field microscopes~\cite{pineChaosThresholdIrreversibility2005,ederaDeformationProfilesMicroscopic2021, sciroccoEffectViscoelasticitySuspending2004,knowltonMicroscopicViewYielding2014,keimMechanicalMicroscopicProperties2014,bowerRheologicalMicrostructuralCharacterisation1999,masschaeleDirectVisualizationYielding2009,hawColloidalGlassesShear1998a}, confocal microscopes~\cite{chanSimpleShearCell2013,linMultiaxisConfocalRheoscope2014,chenMicroscopicStructuralRelaxation2010,shinShearbandingSuperdiffusivityEntangled2017,schallStructuralRearrangementsThat2007,terdikMechanicalTestingColloidal2024,besselingThreeDimensionalImagingColloidal2007,chengImagingMicroscopicStructure2011,c.linBiaxialShearConfined2014}
and magnetic resonance imaging machines \cite{abbottExperimentalObservationsParticle1991,corbettMagneticResonanceImaging1995,raynaudDirectDeterminationNuclear2002,hollingsworthRheonuclearMagneticResonance2004,callaghanRheoNMRShear2008}, as well as in advanced optical setups to measure shear-induced dichroism and birefringence \cite{walesApplicationFlowBirefringence1976,kishbaughRheoopticalStudyShearthickening1993,hongladaromMolecularAlignmentPolymer1993,huKineticsMechanismShear2005,fardinInstabilitiesWormlikeMicelle2012,decruppeFlowBirefringenceStudy1989,lerougeBirefringenceBandingMicellar2004,frattiniDynamicsDiluteColloidal1984,leeSpatialDevelopmentTransient1987}. 
To overcome the tradeoff between sampled volume and spatio-temporal resolution intrinsic to real-space imaging techniques, rheometers have also been coupled to scattering techniques, including small-angle \cite{laugerMeltRheometerIntegrated1995,hashimotoApparatusMeasureSmallAngle1986,huKineticsMechanismShear2005} and wide-angle static light scattering \cite{kushnirWideangleStaticDynamic2021}, homodyne \cite{salmonOpticalFiberBased2003} and heterodyne \cite{salmonLocalRheologyEmulsions2003,giulianiWallVelocimetryRheometer2019} dynamic light scattering, and photon correlation imaging \cite{pommellaCouplingSpaceResolvedDynamic2019}. 
The added value of rheo-scattering experiments is documented by the increasing interest in integrating commercial rheometers in beamlines of large facilities doing small-angle neutron scattering (SANS) \cite{lindnerApparatusInvestigationLiquid1984,cappelaereRheologyBirefringenceSmallangle1997,liberatoreSpatiallyResolvedSmallangle2006,pignonYieldStressThixotropic1997,helgesonRheologySpatiallyResolved2009}, small-angle X ray scattering (SAXS) \cite{panineCombinedRheometrySmallangle2003,denisovSharpSymmetrychangeMarks2015,bauerCollectiveRearrangementOnset2006,dicolaSteadyShearFlow2008,gibaudRheoacousticGelsTuning2020,pignonStructureOrientationDynamics2009,baulandAttractiveCarbonBlack2024}, and X ray photon correlation spectroscopy (XPCS) \cite{lehenyRheoXPCS2015}. 

The integration of shear rheology in the beamline of large scattering facilities benefits from specific geometries compatible with commercial rheometers~\cite{panineCombinedRheometrySmallangle2003}, and no longer requires the development of entirely custom-made instruments as it did a few decades ago~\cite{lindnerApparatusInvestigationLiquid1984,stratyApparatusNeutronscatteringMeasurements1989}, which makes it easily accessible to an increasing scientific community.
Conversely, due to their demanding requirements in terms of optical quality and accessibility, advanced rheo-light scattering experiments are often performed using home-made shear cells. It is the case of many experiments involving Small-Angle Light Scattering \cite{clarkObservationCouplingConcentration1980,salemSmallAngleLight1985,vanegmondTimedependentSmallangleLight1992,sciroccoEffectViscoelasticitySuspending2004,varadanShearInducedMicrostructuralEvolution2001,tamboriniPlasticityColloidalPolycrystal2014,aimeStresscontrolledShearCell2016,aimePowerLawViscoelasticity2018,hawColloidalGlassesShear1998a,pignonButterflyLightScattering1997}, Wide-Angle Light Scattering~\cite{wuEnhancedConcentrationFluctuations1991}, Dynamic Light Scattering \cite{gollubOpticalHeterodyneStudy1974}, Diffusing Wave Spectroscopy \cite{hohlerPeriodicNonlinearBubble1997,hebraudYieldingRearrangementsDisordered1997,petekidisYieldingFlowColloidal2003,kalounAgingColloidalGlass2005,amonHotSpotsAthermal2012,hawColloidalGlassesShear1998a},  and combined imaging and scattering measurements \cite{kumeNewApparatusSimultaneous1995}.
In addition to offering a larger flexibility and easier integration in complex experimental setups, home-made devices provide a wider variety of deformation geometries, not limited to the torsional shear geometry common to commercial rheometers. 
In particular, linear sliding parallel plates are particularly convenient for rheo-optics, because of the homogeneous deformation field combined with the higher quality of flat optical interfaces. In addition, sliding parallel plate geometry can be more easily generalized to 2D shear deformations than torsional rheometers~\cite{linMultiaxisConfocalRheoscope2014,blancRheologyDenseSuspensions2023}.

However, using a home-made shear cell instead of a commercial rheometer comes at a price: the quality of rheological measurements. Indeed, the development of shear cells suitable for soft materials poses significant technical challenges, including minimizing mechanical friction and other spurious contributions to the stress measurement while limiting mechanical vibrations and drifts that would negatively impact optical measurements, and measuring or applying a wide range of mechanical stresses while maintaining sufficient stiffness to ensure precision in gap and applied strain.
As a result among the many shear cells used to deform soft samples in an optical setup, relatively few are actually able to perform rheological measurement. They can be divided in two classes, similar to commercial rheometers: strain-controlled shear cells measure the stress required to apply a prescribed deformation, while stress-controlled ones apply a stress and measure the resulting deformation. The former strategy, exploited since the seminal work of G.I. Taylor \cite{taylorFormationEmulsionsDefinable1934}, typically use stiff piezoelectric~\cite{linMultiaxisConfocalRheoscope2014,terdikMechanicalTestingColloidal2024,wuMicronewtonShearRheometer2022} or stepper motor actuators~\cite{amonHotSpotsAthermal2012, blancRheologyDenseSuspensions2023}, in series with stress transducers, whose compliance sets the sensitivity to small stresses, at the same time limiting the range of large stresses that can be accurately measured without affecting the applied deformation \cite{garritanoCompensatedRheometer1985,garritanoApparatusMethodMeasuring1986,garritanoWideRangeDynamic2005,terdikMechanicalTestingColloidal2024}.
Conversely, stress-controlled shear cells impose the stress via electromagnetic actuators~\cite{aimeStresscontrolledShearCell2016,villaQuantitativeRheomicroscopySoft2022} or using the viscous stress of liquids driven at a controlled shear rate~\cite{vandenbruleSimpleConstantstressRheometer1992,chanSimpleShearCell2013}, and use non-contact position encoders to measure the resulting deformation. This second approach requires to minimize the mechanical friction between moving parts to avoid spurious contributions to the stress reading~\cite{rafferRotaryViscometerAir2001}, which typically makes them more prone to positional drifts requiring specific corrections for microscopy \cite{ederaDeformationProfilesMicroscopic2021,aimeProbingShearinducedRearrangements2019a} and light scattering \cite{aimeProbingShearinducedRearrangements2019}.

The majority of the shear cells presented above is the result of a significant effort in the development of novel strategies to combine actuators, sensors and mechanical components into complex instruments. Building shear cells starting from commercially-available integrated solutions would significantly reduce the development effort, and would open the way to a much wider spreading of rheo-optical experimental platforms. 

In this work, we present a novel two-axes shear cell, composed of one commercially-available air bearing stage, and another prototype air bearing stage specifically designed to optimize its performance for rheology. 
We characterize the air bearing stages, demonstrating that they can be used to perform high-quality shear rheology, including orthogonal superposition rheology experiments. Thanks to its flexibility, this setup can be coupled to a wide variety of techniques. Here, we couple it to both optical microscopy and spatially-resolved light scattering. The resulting rheo-optical setup enables the simultaneous measurement of macro-scale rheology, meso-scale deformation profiles, and micro-scale dynamics in a wide range of soft matter systems. 

This paper is organized as follows: in Sec. \ref{sec:setup}, we present an overview of the whole setup. In Sec. \ref{sec:calib}, we present and calibrate the shear cell actuators, discussing spurious contributions to the mechanical signal that need to be corrected to isolate the contribution of the sample rheology. In addition, we discuss how these actuators can be integrated in a shear cell with carefully-aligned parallel sliding plates. In Sec. \ref{sec:validation}, we validate our shear cell, comparing measurements on a well-characterized soft glassy sample obtained with the shear cell and with a commercial rheometer. In addition, we demonstrate that our shear cell can be used to perform orthogonal superposition rheology. Finally, in Sec. \ref{sec:rheooptics}, we present the optical setup coupled to the shear cell: we use microscopy to demonstrate the absence of wall slip and shear banding, and light scattering to measure microscopic dynamics in the soft glassy sample as it yields. Section \ref{sec:conclusion} concludes the paper with final remarks and potential perspectives.

\section{Setup overview}\label{sec:setup}

\begin{figure}[H]
\centering
\includegraphics[width=\columnwidth]{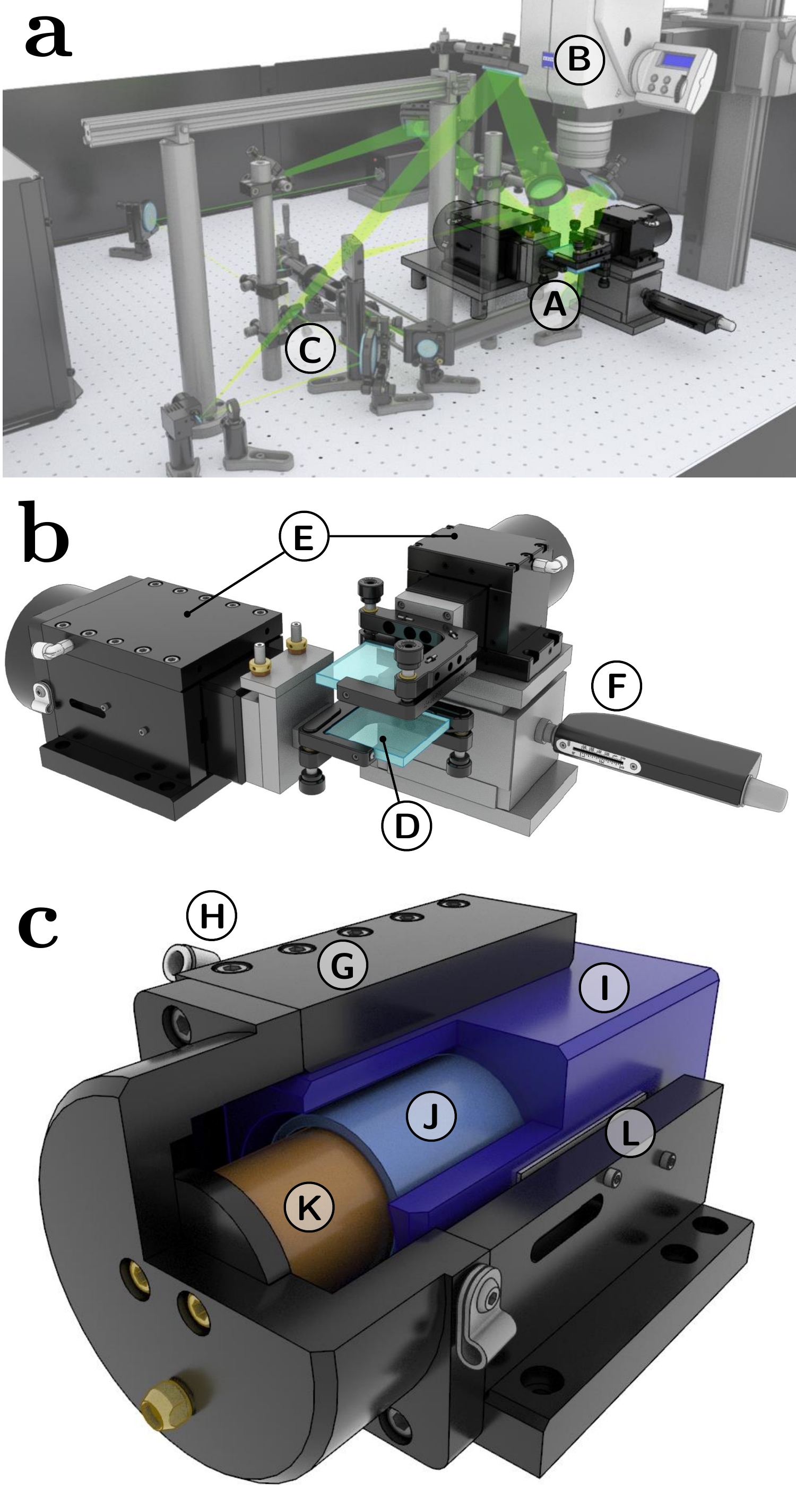}
\caption{\textbf{Setup overview} a) Rendering of the whole rheo-optical setup, coupling the home-made shear cell (A) to an optical microscope (B) and a photon correlation imaging setup (C). b) Sketch of the two-axes shear cell: the sample is confined between transparent parallel plates (D) mounted on linear air bearing stages (E). The gap between the two plates is adjusted using a motorized vertical stage (F). c) Section of the linear air bearing stage, highlighting its main components: air bearing body (G), inlet for compressed air (H), moving carriage (I), voice coil permanent magnet (J) and motor coil (K), position encoder (L). Moving parts are drawn in blue shades for clarity}\label{fig:setup}
\end{figure}

A picture of the setup in its typical configuration is shown in Fig.~\ref{fig:setup}.
The shear cell consists of two parallel plates, 5$\times$5~cm$^2$ in size, confining the sample in a gap that can be controlled by a motorized vertical stage. Both plates are mounted on linear air bearings stages actuated by independent electromagnetic voice coil motors (model A-131 and model A-132, from Physik Instrumente, for bottom and top plates, respectively), as shown in Figure~\ref{fig:setup}b. This allows us to move one plate relative to the other, thereby shearing the sample, with minimal friction, which is crucial for stress measurements~\cite{rafferRotaryViscometerAir2001}. The two air bearing stages are independently mounted on an optical table (M-SST-46-12, from Newport Optics), and the angle $\varphi$ between their translation axes can be adjusted freely between 90$^\circ$ and 180$^\circ$.

With $\varphi=90^\circ$, as shown in Fig.~\ref{fig:setup}, simultaneous actuation of the two axes enables 2D shear protocols, including Orthogonal Superposition Rheology, as described in section~\ref{sec:validation}. With $\varphi=180^\circ$, as shown in Fig.~\ref{fig:parall}a, only 1D shear protocols are available: in this configuration, actuation of both motors can be used to obtain a stagnation plane in between the two plates, convenient for microscopy experiments. Alternatively, one stage (typically, A-131) can be used as strain actuator, while the second (typically, A-132) can be used as force-rebalanced transducer, providing optimized strain-controlled rheology results.

This setup employs voice coil motors, for which the applied force is simply proportional to the electric current provided by the dual-axis servo controller (A-812, from Physik Instrumente). In particular, we use two motors with different strengths: a stronger one, mounted on A-131, able to apply forces up to 20~N, and a weaker one, mounted on A-132, with 10-times higher sensitivity to smaller forces, below 1~mN. Considering a typical cross-sectional area of a few cm$^2$, this translates in the ability to apply or measure stresses ranging between about 1 and $10^4$~Pa. 
Linear encoders installed on the two air bearing stages and read simultaneously by the controller every 0.5~ms measure the displacement of both plates with a nominal precision of 1.2~nm. This displacement is then used to compute the shear deformation: by relating the imposed stress to the measured deformation, we obtain stress-controlled rheological measurements. 
For strain-controlled measurements, the motor controller implements a servo loop running at 20~kHz, allowing us to control the plate position with a precision below 100~nm, corresponding to a precision on the applied strain of about 0.01~\%, and convert the current applied by the motor into a stress measurement, as detailed in Section~\ref{sec:calib}.
The relative orientation of the two translation axes can be adjusted at will, to achieve both simple uniaxial shear and 2D shear deformations.

To detect microscopic dynamics under shear and resolve them in space and time, the shear cell is coupled to a photon correlation imaging (PCI) setup. The sample is illuminated with an expanded laser beam, and light scattered at a well-defined angle $\theta$ is filtered by a diaphragm, collected by a lens and directed to a CMOS camera, which records an image of the entire sample being sheared. When the diaphragm aperture is reduced, the sample image turns into a speckle pattern, each speckle collecting light scattered by particles in a tiny scattering volume, localized in space. 
We analyze the temporal fluctuations of the speckle intensity to measure microscopic dynamics, as detailed in section \ref{sec:PCI}.

Finally, an optical microscope (ZEISS AxioZoom) is also integrated into the setup. It uses a zoom lens to image the sample onto a second CMOS camera. The magnification can be adjusted between 0.2$\times$ and 3.5$\times$, with a constant working distance of about 12.5~cm. This unusually large working distance allows PCI and microscopy to be performed simultaneously, though it introduces trade-offs in magnification and numerical aperture (NA). The NA of the objective can be adjusted between about 0.05 and 0.125 using a diaphragm, enabling different optical measurements under the same laser illumination used for PCI. With the diaphragm closed to minimize the numerical aperture of the AxioZoom, we image a speckle pattern similar to the one previously described, carrying information on a wide scattering angle, supplementary to the one probed by the PCI setup, to probe motion along the shear gradient direction~\cite{aimeDynamicSpeckleHolography2021}. 
When the diaphragm is open, we capture high-resolution images of the sample, with a depth of focus as low as $70~\mu$m at the largest magnification. This can be used to monitor shear-induced structural changes, or to reconstruct the flow profile across the sample gap using Particle Tracking Velocimetry (PTV). To this end, the AxioZoom is mounted on a motorized stage controlled by the PC, which enables scanning of the sample gap with a resolution of about 1~$\mu$m, and can be equipped with a notch filter (NF533-17, from Thorlabs), which blocks scattered laser light and allows imaging of fluorescent tracer particles, as detailed in section~\ref{sec:PTV}.

\section{Shear cell calibration}\label{sec:calib}

The first step in developing a shear cell is the mechanical calibration of the moving stage and force actuator. A section of the air bearing stage highlighting its main components is shown in Fig.~\ref{fig:setup}c. 
The moving carriage holding the shear cell plate is hosted by an air bearing and confined on four sides by compressed air. This allows for its quasi-frictionless sliding along the third axis, pushed by a voice coil motor composed of a fixed coil inserted in a permanent magnet fixed to the carriage. 

The current supplied to the voice coils motors is driven by the servo controller, whose firmware runs a custom-made code that records both carriage positions, $x$, read by the linear encoders, and the current flowing in the voice coils, $I$. 
To extract rheological measurements from raw current and position data, several parameters need to be calibrated and optimized.
First, strain-controlled rheology requires a calibration of the servo loop parameters to optimize positioning accuracy and noise on the force measurement (Sec.~\ref{sec:servo}).
Then, we need to measure the conversion factor relating $I$ to the force applied by the motor coil, $F(I)$ (Sec.~\ref{sec:current}). 
Next, we need to develop a mechanical model describing the inertia and frictional forces acting on the air bearing, to extract the force applied to the sample itself, filtering out other spurious mechanical or electromagnetic contributions (Sec.~\ref{sec:viscous} and \ref{sec:static}).
This mechanical calibration highlights the range of viscoelastic moduli that can be measured with this shear cell (Sec.~\ref{sec:calibres}). 
Finally, we show how to measure the sample cross sectional area and gap height, needed to convert force and displacement data to stress and strain (Sec.~\ref{sec:geometry}).

\subsection{\label{sec:servo}Servo loop optimization for strain-controlled rheology}

The servo controller comes with built-in control loops for operating the air bearing stages in position-controlled mode. The control loop scheme has a nested  structure, including a current loop used to control the current flowing in the voice coil, as well as two loops for position and velocity control, which are active when the stage is operated in strain-controlled mode (so-called closed-loop operation), and are deactivated when the stage is operated in force-controlled mode (open-loop).

Several parameters need to be optimized to achieve optimal controller performance: three coefficients tuning the proportional (P), integral (I) and derivative (D) corrections typical of standard PID controllers, as well as cutoff frequency and strength of bandpass filters embedded in each servo loop to improve stability. The controller comes with a set of factory-configured parameters, chosen to maximize the position accuracy of the bare air bearing.
For optimal performance, these parameters need to be fine-tuned once the additional mass of the shear cell plates and alignment mechanical components is fixed to the air bearing carriage. 
In addition, position accuracy comes at the expense of force measurement in closed-loop operation, as a tiny mismatch between actual and prescribed positions triggers a rapid and strong response of the controller. To improve force accuracy, we slightly detuned the servo controller, reducing its responsiveness to positional errors and lowering its characteristic response frequencies to 170~Hz and 34~Hz for the two motors, respectively. 
These frequencies limit the range of fast timescales that can be probed using this shear cell. In this detuned configuration, we characterized the typical position error, defined as the mean squared difference between actual and prescribed position, for oscillatory motion at various frequencies and amplitudes. The result highlights that the shear cell can be operated at frequencies up to 100~rad/s with a position accuracy better than 0.1~$\mu$m. 

\subsection{\label{sec:current}Inertia and current-force conversion}

\begin{figure}[ht]
    \centering
    \includegraphics[width=\columnwidth]{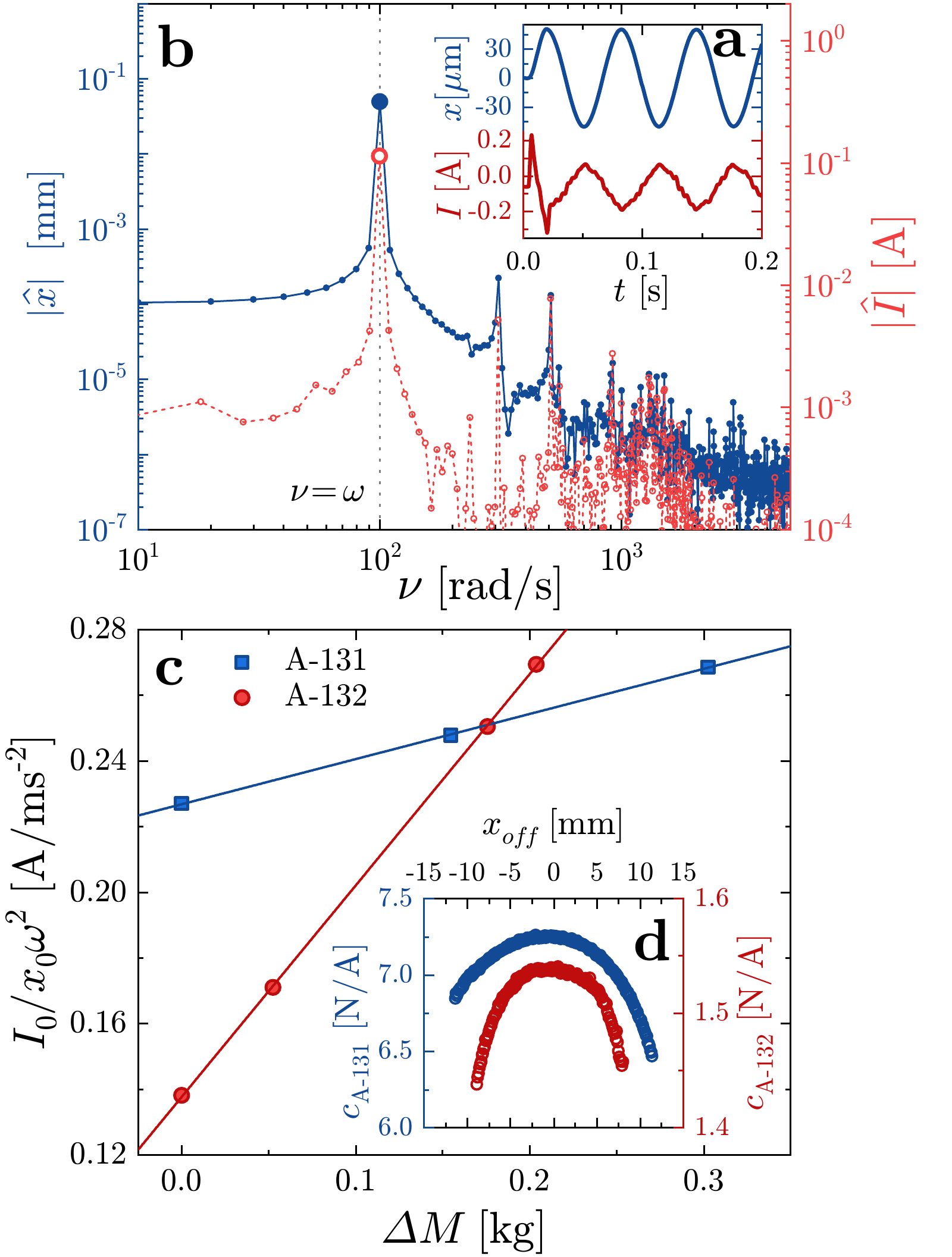}
    \caption{\textbf{High-frequency oscillations} a) oscillatory motion with frequency $\omega=100$~rad/s and amplitude $x_{0}=0.05$~mm, started at time $t=0$. Raw data of position (blue) and current (red). b) Amplitude of the Fourier spectrum for $x(t)$ (blue, left axis) and $I(t)$ (red, right axis). Large symbols and dashed gray line: applied frequency, $\nu=\omega$.
    c) Symbols: ratio between current and acceleration as a function of the added mass, $\Delta M$, for motor A-131 (blue squares) and A-132 (red circles), measured for $\omega=100$~rad/s. Solid lines: linear fit of $I_0/x_0\omega^2=(M+\Delta M)/c$, yielding the carriage mass, $M$, and the current-force conversion coefficient, $c$. d) Current-force conversion factor for motor A-131 (blue, left axis) and A-132 (red, right axis), as a function of carriage position, $\bar{x}$ (increasing as the carriage moves out of the air bearing).}
    \label{fig:rawdatacurrent}
\end{figure}

To characterize the force-current conversion of the air bearing, as well as the inertia of its moving mass, we impose an oscillatory motion of the bare air bearing stages, $x(t)=\bar{x}+x_0e^{i\omega t}$, around an offset position $\bar{x}$, and we measure the current, $I(t)$, required to achieve it. 
One representative result for the first stage (A-131), with $\omega=100$~rad/s and $x_0=50~\mu$m, is shown in Figure~\ref{fig:rawdatacurrent}a. We find that after an initial current spike applied to accelerate the moving mass of the carriage, initially at rest, $I(t)$ exhibits oscillations around a non-zero value, $\bar{I}=3$~mA, which we attribute to gravity, assuming that the sliding bar is not exactly horizontal. 
Horizontal alignment is challenging because of the pneumatic vibration isolators (S-2000, from Newport Optics) supporting the optical table and isolating the setup from building vibrations. The automatic re-leveling system that adapts the air pressure in the isolators in response to a change in the load distribution on the table has a precision of about 1~mrad, resulting in spurious constant contributions to $\bar{I}$ due to gravity, that we calibrate before each experiment by measuring the current required to keep the carriage motionless.
To characterize the current oscillation shown in Figure~\ref{fig:rawdatacurrent}a, we analyze $I(t)$ in the Fourier domain, by selecting a time window of an integer number, $n=10$, of oscillation periods, excluding the initial and final transient accelerations, and computing its frequency spectrum, $\hat{I}(\nu)$. The result is discretized in units of $\nu_{min}=\omega/n$, the smallest frequency accessible to the analysis. We find that both $|\hat{x}|$ and $|\hat{I}|$ peak at the $n$-th datapoint, at $\nu=\omega$, corresponding to the first harmonic amplitudes $\hat{I}_0=\hat{I}(\omega)$ and $\hat{x}_0=\hat{x}(\omega)$.

We find that $|\hat{x}_0|=0.05$~mm, equal to the imposed oscillatory amplitude, $x_0$, and $|\hat{I}_0|=0.11$~A. We interpret the current amplitude assuming that, in absence of other mechanical forces acting on the air bearing, $F$ is directly proportional to the acceleration of the moving carriage, $\ddot{x}=-x_0\omega^2e^{i\omega t}$, such that its first-harmonic amplitude is $\hat{F}_0=\hat{F}(\nu=\omega)=-M\omega^2\hat{x}_0$, with $M$ the moving mass of the air bearing carriage, which for the A-131 stage is nominally $M=1630$~g. Assuming a linear relationship between force and current, $F=cI$, we use the measured values of $\hat{I}_0$ and $\hat{x}_0$ to extract $c=M\omega^2|\hat{x}_0/\hat{I}_0|=7$~N/A, in agreement with the nominal value. 

To better characterize the force/current conversion factor, we repeat this measurement for the same value of $\omega$ and different oscillatory amplitudes $x_0$, obtaining the expected linear relationship, 
from which we determine the ratio $M/c$. To measure independently $M$ and $c$ without relying on the nominal values, we repeat the same experiment by changing the moving mass by well-controlled amounts, $\Delta M$. For all moving masses we obtain a linear relationship, and we find that the slope increases linearly with $\Delta M$, as shown in Figure~\ref{fig:rawdatacurrent}c. From this increase we obtain refined values for the current-force conversion factor $c_1=7.2$~N/A, and for the carriage mass, $M_1=1.65$~kg. 
Repeating the same experiment for the second air bearing stage, A-132, we find $c_2=1.563$~N/A and $M_2=0.215$~kg. Both values are substantially reduced with respected to A-131: indeed, the A-132 air bearing stage is a prototype model, designed upon our request to minimize the contribution from inertia and to maximize the sensitivity to smaller stresses.

Because the electromotive force of the voice coil depends on the coil position in the spatially-heterogeneous magnetic field of the permanent magnet, the current-force conversion factor $c$ is susceptible to slightly change as the carriage moves. To characterize this change, we repeat these oscillatory tests for different offset positions, $\bar{x}$, spanning the full accessible range. We find that for both motors $c$ has a bell shape, decaying by about 10\% moving towards the edges of the allowed displacement, as shown in Figure~\ref{fig:rawdatacurrent}d. 
Based on this result, we choose to perform our experiments starting from the offset position where $c(\bar{x})$ is maximum, and where its spatial variations are negligible.

\subsection{\label{sec:viscous}Force noise and counter-electromotive force}

To investigate other potential factors affecting our rheological measurements, we study the impact of the oscillatory frequency, by repeating the calibration described above for smaller values of $\omega$. For each $\omega$, we vary the amplitude, $x_0$, up to a maximum $x_{0,Max}=\textrm{min}(F_{Max}/M\omega^2, x_{0,th})$, set by the maximum force exerted by the voice coil, $F_{Max}$, and by an arbitrary threshold $x_{0,th}=5$~mm, chosen to limit spatial variations of $c(x)$ to less than 2\%, so that they will be neglected in the following. Considering a typical sample thickness $h_{typ}\sim 0.3-1$~mm, this threshold corresponds to about 10 strain units.

\begin{figure}[ht]
    \centering
    \includegraphics[width=\columnwidth]{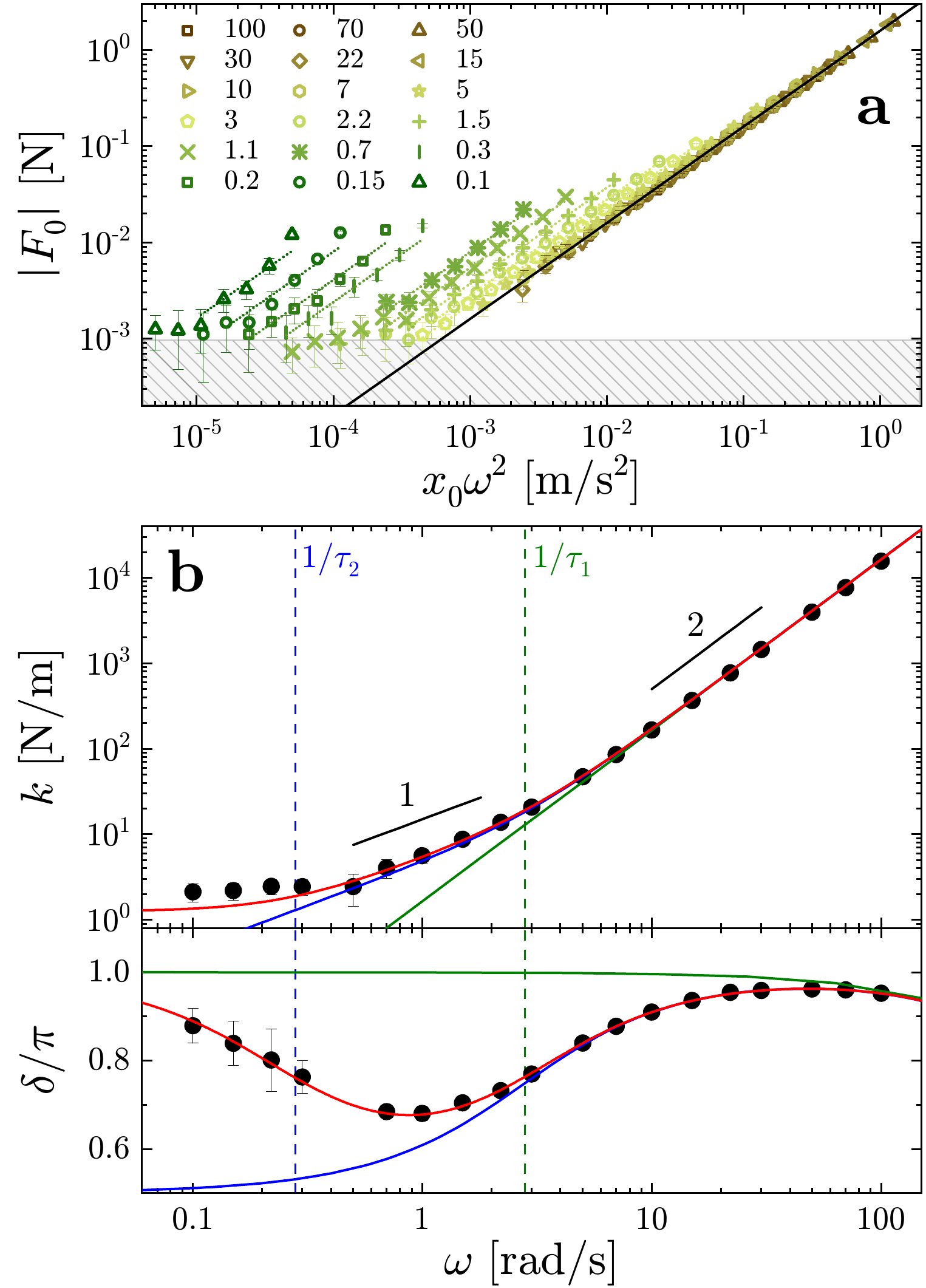} 
    \caption[Fvsa0]{\textbf{Frequency dependence} a) Force amplitude as a function of the imposed acceleration amplitude, measured on the A-131 stage for different angular frequencies, increasing from 0.1~rad/s (green) to 100~rad/s (brown), as specified in the legend. Black solid line: $|\hat{F}_0|=Mx_0\omega^2$, with $M=1.65$~kg the carriage mass. Gray shaded area: $|\hat{F}_0|<F_{noise}$ with force noise level $F_{noise}=1$~mN. Dashed lines: linear fits for $|\hat{F}_0|>F_{noise}$
    b) Amplitude (top) and phase (bottom) of mechanical susceptibility $\hat{F}_0/\hat{x}_0$. Black symbols: experimental datapoints extracted from linear fits in panel a. Green solid line: pure inertial model with $M=1.65$~kg and $t_f=1.2$~ms. Blue solid line: visco-inertial model with same parameters as above and $\lambda=4.3$~Ns/m. Red solid line: full model with same parameters as above and $p=-1.3$~N/m.}\label{fig:Fvsa0}
\end{figure}

To highlight deviations from the inertial regime discussed in section~\ref{sec:current}, we plot the amplitude of the force oscillations, $|\hat{F}_0|$, as a function of carriage acceleration, $x_0\omega^2$. 
If inertia was the only contribution to the mechanical force, all curves should collapse on a straight line with slope $M$. This is indeed the case for the highest angular frequencies, $\omega > 3$~rad/s, whereas for lower values of $\omega$ the expected linear trend underestimates the measured force, suggesting the presence of additional mechanical contributions, as shown in Figure~\ref{fig:Fvsa0}a.

Our calibration data highlights two sources of deviation from the expected inertial regime. The first one is represented by a frequency-independent threshold force value of about $F_{noise}=1$~mN, which we attribute to noise: when $|\hat{F}_0|\approx F_{noise}$, the first-harmonic peak in the force spectrum, $\hat{F}(\nu)=c\hat{I}(\nu)$, becomes comparable with the baseline noise value, $I_{noise}\approx 0.2$~mA, well-visible at high frequency as shown in Figure~\ref{fig:rawdatacurrent}b.
For $|\hat{F}_0|> F_{noise}$, we find that $\hat{F}_0$ grows linearly with $\hat{x}_0$ for all values of $\omega$: this allows for the definition of a mechanical susceptibility, $k=|\hat{F}_0/\hat{x}_0|$, that is shown in Figure~\ref{fig:Fvsa0}b along with the phase delay, $\delta=\arg(\hat{F}_0/\hat{x}_0)$, as a function of $\omega$.

In the regime of high frequencies, we find that $k \propto \omega^2$, and $\delta\approx \pi$. This is the regime where the mechanical response is dominated by inertia. 
As $\omega$ decreases, we observe that $k$ deviates upwards, indicating an additional contribution to the applied force. At the same time, $\delta$ decreases, suggesting that this additional contribution is viscous-like. We characterize its magnitude by observing that $k\propto \omega$ in an extended frequency range, $0.4\leq\omega\leq 4$~rad/s, from which we extract $k/\omega \approx 5$~Ns/m. 
This value is much larger than the expectations based on the viscous friction of the air film supporting the bearing carriage: considering a total film surface of $S_{air}=$0.03~m$^2$, an average film thickness of $h_{air}=10~\mu$m and a dynamic viscosity of air at operating pressure (550~kPa) of about $\eta_{air}=2~\mu$Pa$\cdot$s, we obtain $k/\omega \sim \eta_{air}S_{air}/h_{air}\sim 6\cdot 10^{-3}$~Ns/m, about 1000 times smaller than the observed value. 

\begin{figure}[H]
    \centering
    \includegraphics[width=\columnwidth]{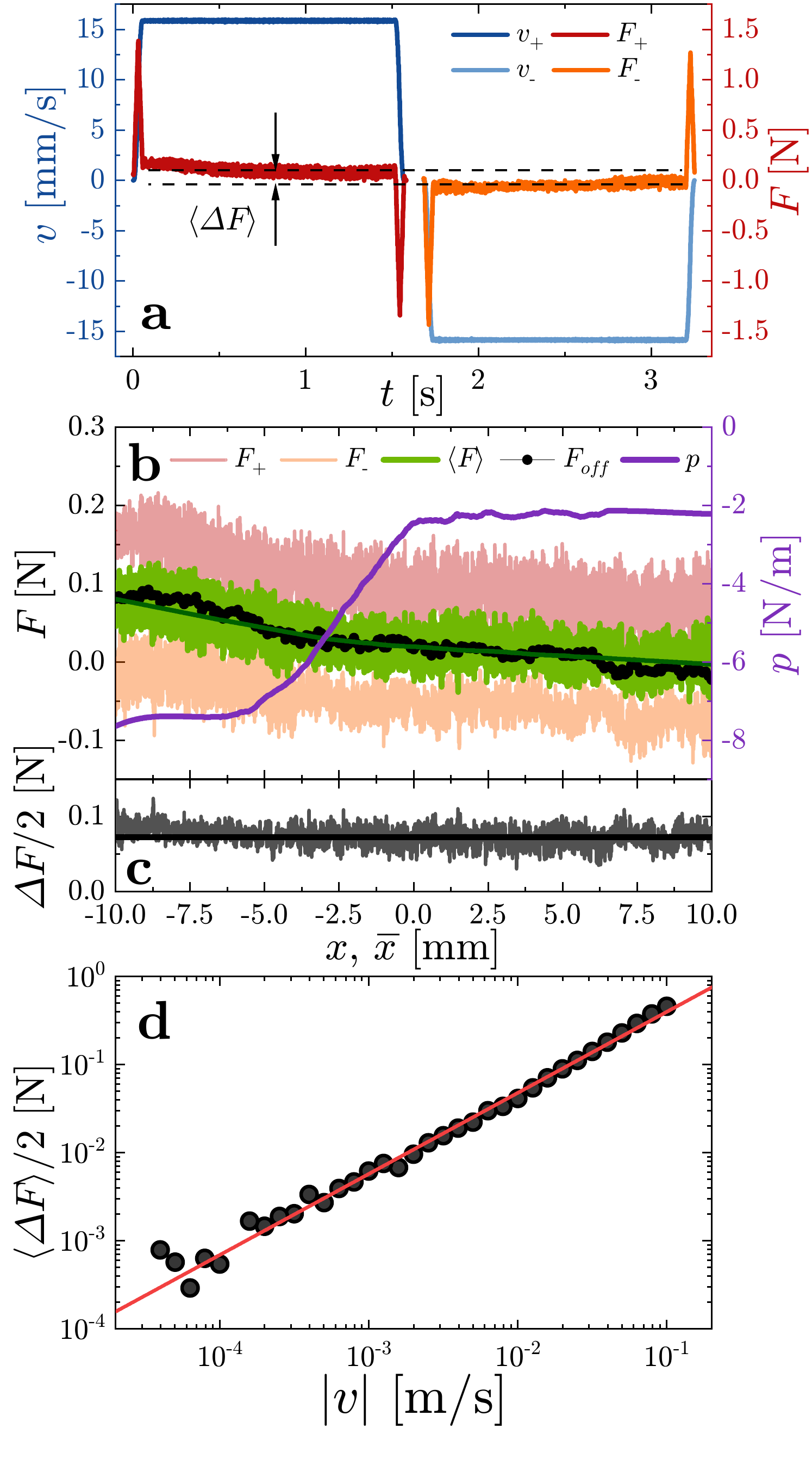} 
    \caption{\textbf{Constant speed} (a) Speed (blue, left axis) and force (red, right axis) measured on the A-131 stage for an imposed $|v|=15$~mm/s; (b) left axis: force during ascending (red) and descending (orange) ramp, as a function of carriage position, $x$, excluding initial and final accelerations. Green line: averaged force, $\langle F\rangle=(F_++F_-)/2$. Dark green line: smoothed $\langle F\rangle$. Black symbols: static offset force during oscillatory measurements at offset position $\bar{x}$. Purple line, right axis: spatial derivative of smoothed $\langle F\rangle$. (c) Gray line: steady-state force value corrected for static force. Black solid line: average on all carriage positions. (d) Black symbols: average force measured under constant speed $v$. Red line: linear fit yielding $\lambda=4.3$~kg/s.}\label{fig:emf}
\end{figure}

Therefore, we attribute this effect to the counter-electromotive force (back EMF) generated by the relative motion of magnet and coil \cite{tiplerPhysics1982}. 
Following Faraday's law of induction, the motion of the magnet induces a current in the coil, which is proportional to the carriage speed, and opposite in sign to the current required to produce that specific motion. 
Since the physical mechanism responsible for the back EMF is the same as that generating the electromotive force itself, the voltage induced in the coil per unit speed is described by the same constant, $c_1\approx 7~N/A=7~V\cdot s/m$, and the current induced per unit speed will be $c/R$, with $R$ the coil resistance. The back EMF will then be $I_b=-v\cdot c/R$, resulting in an apparent viscous drag force, $F_v=-\lambda \dot{x}$, with $\lambda=c^2/R$ representing an effective drag coefficient. Under oscillatory motion, this drag manifests as a force response linear in $\omega$ and characterized by a $\pi/2$ phase delay relative to the position: $\hat{F_v}=-i\lambda\omega\hat{x}$. This prediction is in line with the experimental results shown in Figure~\ref{fig:Fvsa0}b, where $\delta$ decreases as $\omega$ is decreased below 10~rad/s. 

To better characterize $F_v$ independently of inertia effects, we perform a different series of experiments where the sliding bar of the shear cell is moved at constant speed, $v$. One representative result is shown in Figure~\ref{fig:emf}a: starting from a quiescent condition, $v$ is rapidly raised to the prescribed value, and is subsequently kept constant until motion is stopped close to the maximum attainable displacement. The initial acceleration and the final deceleration are associated to peaks in the applied force due to inertia. Away from these peaks, the force is small and nearly constant. In this regime, inertia does not contribute, and the force signal is dominated by the back EMF, therefore it can be used to characterize the effective drag coefficient, $\lambda$.
To extract $\lambda$ independently from the residual static contribution, we repeat the experiment, moving the carriage in the opposite direction. We find that the two force profiles have similar shape, but are shifted by an amount $\Delta F=2\lambda v$. We measure $\Delta F$ by taking the difference of the two force signals at the same offset positions, and then averaging the difference on all available positions. The resulting measurement of $\langle \Delta F\rangle$ grows linearly with $v$, across more than three decades, as shown in Figure~\ref{fig:emf}d. From a linear fit we obtain $\lambda_1=4.3$~kg/s, corresponding to a coil resistance $R_1=c_1^2/\lambda_1=11~\Omega$. A similar analysis for the second stage yields $\lambda_2=0.125$~kg/s, corresponding to a resistance $R_2=18~\Omega$.

We integrate this apparent force in the equation of motion for the sliding bar, which becomes $F=M\ddot{x}-\lambda \dot{x}$, yielding $k=-M\omega^2-i\lambda\omega$ under oscillatory motion at angular frequency $\omega$. The magnitude of the mechanical susceptibility is then:

\begin{eqnarray}
\label{eqn:amplitude}
     |k|=M\omega^2 \sqrt{1+\frac{1}{\tau_1^2\omega^2}} \\
     \tan(\delta) = \frac{1}{\omega \tau_1} 
\end{eqnarray}

where $\tau_1=M/\lambda=0.39$~s and 1.72~s for A-131 and A-132, respectively, is a characteristic time capturing the crossover between inertia-dominated and back EMF dominated regimes in Figure~\ref{fig:Fvsa0}b.
  
\subsection{\label{sec:static}Static force}

At even lower frequencies, we find that $k(\omega)$ flattens out to a plateau value, whereas $\delta$ increases again, deviating from the prediction of the model based on back EMF and inertia.
To model this additional deviation, we observe that the force shown in Figure~\ref{fig:emf} shows a small but measurable dependence on $x$. The origin of this dependence is still unknown: we speculate that it might be due to either the imperfect flatness of the carriage or to slight variations in the pressurized air flux corresponding to different carriage positions. 
Building upon this empirical observation, we use the force measured under very small $v$, we average on both carriage directions to compute a position-dependent offset force, $\bar{F}(x)$, and we study its impact on oscillatory motion. We observe that spatial gradients in $\bar{F}$ result in a time-dependent contribution to the force measured as the carriage position oscillates. In particular, a positive local gradient, $p=\partial \bar{F}/\partial x>0$, increases the instantaneous force measured when $x>\bar{x}$, and decreases the  instantaneous force measured when $x<\bar{x}$. As such, it acts as an elastic recoil. Conversely, a negative gradient, $p<0$, would act as an augmented inertia term. Assuming that $p$ is approximately constant in the range of positions explored by the imposed motion profile, the full model reads: $F=M\ddot{x}-\lambda \dot{x}+px$, and under oscillatory motion it yields:

\begin{equation}
\label{eqn:amp_full}
     |k|= \sqrt{(M\omega^2-p)^2+\lambda^2\omega^2}
\end{equation}

\begin{equation}
\label{eqn:phase_full}
     \tan(\delta) = \frac{1}{\omega \tau_1-\frac{1}{\omega\tau_2}} 
\end{equation}

where $\tau_2=\lambda/|p|$ represents a second characteristic time, setting a second crossover to a third, low-frequency regime where the measured mechanical properties are dominated by local variations in $\bar{F}$. 
Fitting the frequency-dependent stiffness from Figure \ref{fig:Fvsa0}b, we obtain $\tau_2=3.7$~s, corresponding to $|p|\approx 1$~N/m, in fair agreement with $p\approx -2$~N/m estimated from Figure \ref{fig:emf}b.

To complete our mechanical model, we observe that, at the highest frequencies, force and position have a relative phase difference $\delta\approx 3$~rad, slightly lower than the expected value $\delta=\pi$, and decreasing with increasing $\omega$. This effect is due to the slightly asynchronous reading of force and position: to improve the accuracy of force measurements, the measured value of $I(t)$ is the result of a moving average performed over the previous $N=4$ controller cycles, which effectively introduces a relative delay $t_f \approx 1$~ms between position and force readings, such that the effective phase delay in the inertial regime becomes $\delta=\pi-\omega t_f$.

\subsection{\label{sec:calibres}Mechanical calibration results}

In conclusion, the mechanical calibration of our two air bearing stages yielded a mechanical model including: (1) a moving mass, $M$, responsible for inertia dominating the mechanical signal at the highest frequencies; (2) the voice coil electromotive force (EMF), quantified by a force/current conversion factor, $c$, and by the internal resistance, $R$, dictating the strength of the back EMF that plays the role of an effective viscous drag, $\lambda$; (3) a position-dependent static force offset, whose spatial gradient, $p$, enters the constitutive equation as an effective elastic recoil term, which dominates at the lowest frequencies.
The interplay between these three effects can be described in terms of two characteristic timescales, $\tau_1=M/\lambda$ and $\tau_2=\lambda/|p|$. The values of the calibration parameters and of the two timescales are summarized in Table \ref{tab:param}.

\begin{table}[h]
\caption{Air bearing stage mechanical calibration parameters}\label{tab:param}%
\begin{tabular}{@{}lllll@{}}
Symbol & AX0 & AX1 & Unit & Description \\
\midrule
$x_{max}$          & 25   & 20    & mm       & maximum travel\\
$c=F/I$            & 7.2  & 1.5   & N/A      & $F(I)$ conversion\\
$M$                & 1.65 & 0.215 & kg       & Moving mass\\
$t_f$              & 1.2  & 1.2   & ms       & Delay time\\
$\lambda$          & 4.3  & 0.125 & kg/s     & Back EMF drag\\
$R=c^2/\lambda$    & 11   & 18    & $\Omega$ & Coil resistance \\
$p$                & -2    & 1    & N/m      & $\bar{F}$ gradient (typ.)\\
$\tau_1=M/\lambda$ & 0.39 & 1.72  & s        & Inertial to viscous\\
$\tau_2=\lambda/|p|$ & 3.7  & 0.1 & s        & Viscous to static\\
$F_{noise}$        & 1     & 1    & mN       & $F$ noise (typ.)\\
\botrule
\end{tabular}
\end{table}

To better illustrate the impact of the model parameters for the performance of the linear actuator, we model the outcome of the oscillatory rheology measurement of two classes of ideal samples: pure Hookean solids with elastic modulus $G$, and Newtonian liquids with shear viscosity $\eta$. We assume that the sample cross-sectional area is $S=4$~cm$^2$, and its thickness is $h=0.5$~mm. 
We use $S$ and $h$ to convert the sample stress into a mechanical force, $F_S(x, \dot{x})$, with $x$ and $\dot{x}$ the position and speed of the shear cell plate. This force adds to the mechanical model, such that: $F=M\ddot{x}-\lambda \dot{x}+px+F_S$. In terms of complex amplitudes: $\hat{F}_0/\hat{x}_0=p-M\omega^2-i\omega\lambda + \hat{F}_{S,0}/\hat{x}_0$, with $\hat{F}_{S,0}/\hat{x}_0=GS/h$ for elastic samples and $i\omega\eta S/h$ for viscous samples.

\begin{figure}[ht]
    \centering
    \includegraphics[width=\columnwidth]{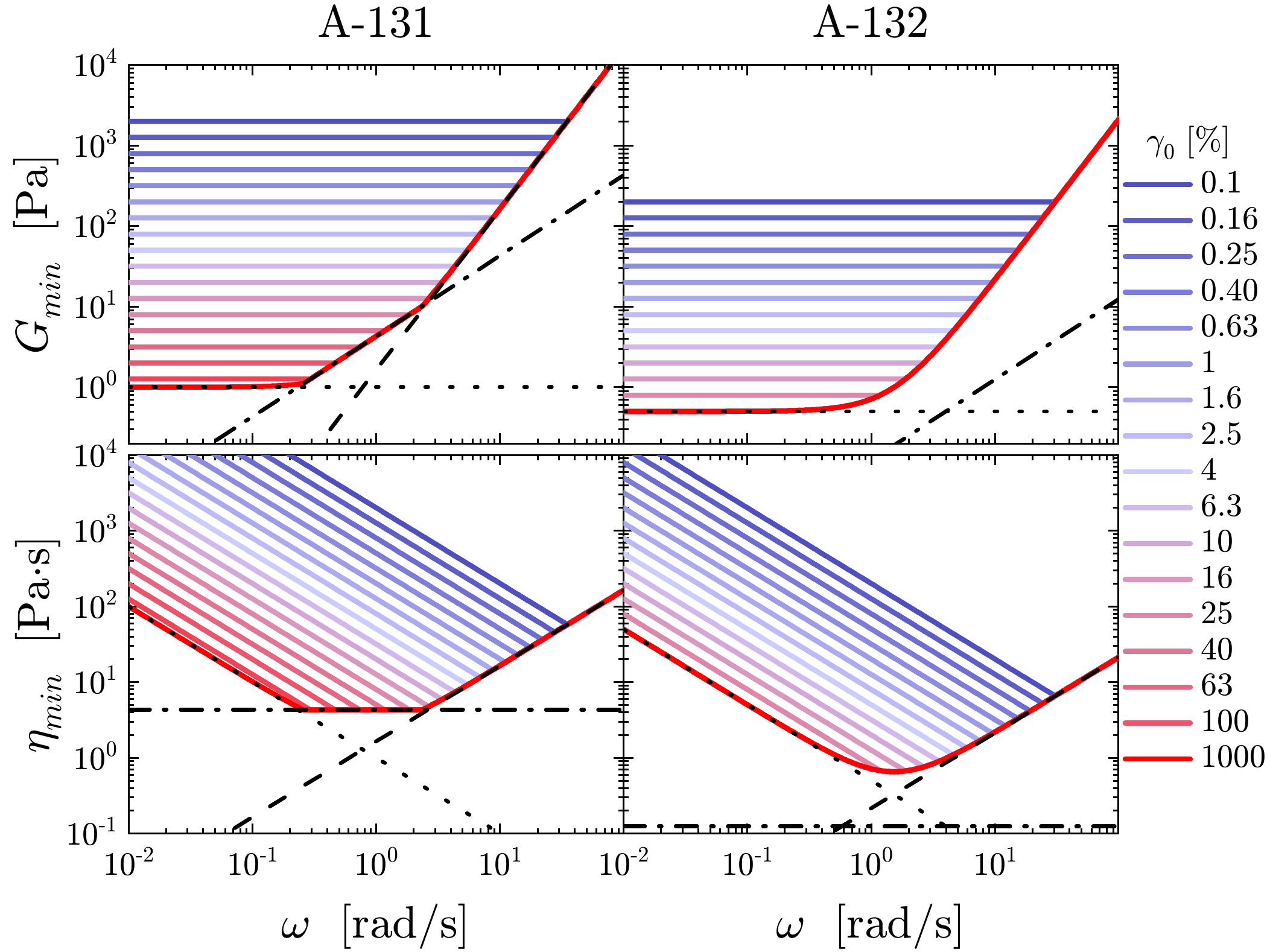} 
    \caption{\textbf{Calibration result: measurable samples} Solid lines: minimum values of $G$ (top row) and $\eta$ (bottom row) measurable under oscillatory shear with air bearing stage A-131 (left column) and A-132 (right column). Color shades from blue to red represent increasingly large strain amplitudes, as reported in the label. Black dashed, dash-dotted and dotted lines correspond to the limits set by inertia, back EMF and static force gradients, respectively.}\label{fig:regimes}
\end{figure}

For elastic samples, $F_S$ dominates the mechanical signal in a range of frequencies such that $GS/h>|M\omega^2-p|$, and that $GS/h>\omega\lambda$, and in a range of strain amplitude limited by noise: $\gamma_0=|\hat{x}_0|/h>F_{noise}/GS$. This gives a conservative estimate of the minimum elastic modulus, $G$, that can be properly measured at a given $\omega$ and $\gamma_0$. A similar argument can be used to define, the minimum value of $\eta$ that can be measured in a viscous sample. These values are reported in Figure \ref{fig:regimes} as a function of $\omega$, for selected values of $\gamma_0$. 

Figure \ref{fig:regimes} highlights that the air bearing stages can be exploited to measure the oscillatory rheology of soft solids with yield strain larger than 1\% and elastic moduli larger than 10~Pa, which encompass a wide range of materials of interest. 
By contrast, we expect our shear cell to be less suitable to probe the oscillatory rheology of liquid-like samples with viscosities below $10$~Pa$\cdot$s. For viscous materials, we expect constant-rate experiments to be more suitable: they are unaffected by inertia, and the contribution of static force gradients is corrected by reverting the carriage direction, such that the only leftover contribution, other than mechanical noise, will be that of back EMF, setting a threshold $\eta_{min,\dot\gamma}\approx 0.1$~Pa$\cdot$s for A-132. 
If needed, this value can be further reduced by increasing the coil resistance, for instance by adding an external resistance in series, which should proportionally reduce the back EMF and increase sensitivity to viscous stresses.

\begin{figure}[H]
    \centering
    \includegraphics[width=\columnwidth]{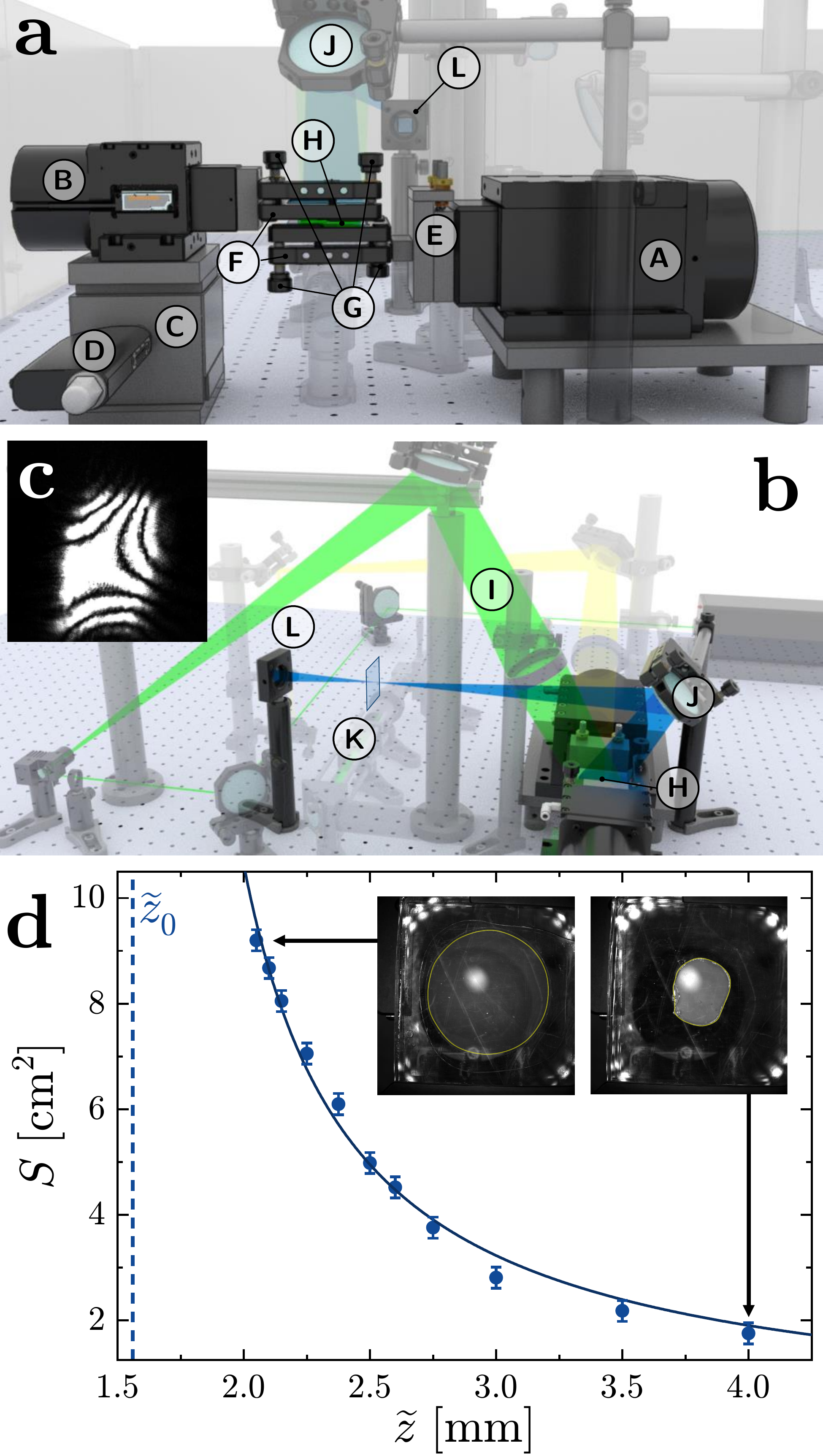}
    \caption{\textbf{Plate alignment} a) Sketch of two air bearing stages A-131 (A) and A-132 (B) integrated in the shear cell. (C) vertical translation platform; (D) linear stepper motor; (E) adjustable stage to set the vertical position of the lower plate; (F) kinematic mounts for the parallel plates; (G) micrometric screws to adjust plate parallelism and orientation relative to the translation axes; (H) sample; (J,L) optics for parallelism adjustment. Inessential optical elements have been rendered as partially transparent to improve the clarity of the sketch.
    b) Sketch of the optical setup for plate alignment. (I) incoming (green) and partially reflected (blue) laser beams. The different color is for visualization purpose only. (J) mirror collecting the reflected light, and sending it to a CMOS camera (L); (K) beam waist position, where a screen is added for coarse alignment (sketched as a blue rectangle). c) representative interference pattern recorded by the CMOS camera when the two plates are well aligned. d) Symbols: sample surface as a function of nominal position of the vertical translation stage, $\tilde{z}$. Line: fit, extracting the nominal contact position, $\tilde{z}_0$. Sample thickness will then be $h=\tilde{z}-\tilde{z}_0$. Insets: snapshots of the sample while lowering the top plate. The sample interface has been highlighted in yellow for clarity.}\label{fig:parall}
\end{figure}

Alternatively, clean oscillatory rheology measurements without spurious mechanical contributions arising from the carriage motion can be obtained by taking advantage of the double motor: by aligning the two air bearing stages as shown in Figure \ref{fig:parall}a, such that the two translation axes are parallel, we can use one stage (typically, A-131) as strain actuator, and the other stage as a force-rebalanced transducer~\cite{garritanoCompensatedRheometer1985}: in this configuration, $F_{noise}$ is the only limiting factor setting the range of measurable stresses to $\sigma_{noise}=F_{noise}/S\sim 1$~Pa.

\subsection{Measuring gap and cross-sectional area}\label{sec:geometry}

To use our air bearing stages to measure shear rheology, we mount two flat, transparent plates to both air bearing carriages, as shown in Figure~\ref{fig:setup}b. The two plates, 5x5~cm in size, are held by kinematic mounts (KM200S, from Thorlabs), allowing us to finely control their alignment. Their transparency ensures optical access to the sample for PCI and microscopy, and enables fine-tuning their parallelism through interferometry, as discussed hereafter.
To adjust the distance between the two plates, we mount the A-132 stage on a motorized vertical translation platform (M-MVN80 with linear stepper motor LTA-HL, from Newport Optics), which has a maximum travel range of 12.5~mm and a positioning accuracy of about $0.5~\mu$m. In addition, the overall vertical position of the sample can be adjusted thanks to micrometric screws (DAS110, from Thorlabs) coupled to a home-made mounting bracket.

Aligning the two plates before loading the sample is critical to ensure a homogeneous gap and thus a homogeneous deformation, which is key to properly measure rheological properties, especially in the nonlinear regime. Commercial rheometers have a typical alignment precision below 1~mrad, limiting spatial variations of the imposed deformation field below 1\% \cite{rodriguez-lopez_using_2013}. To achieve a similar precision in our shear cell, we use an interferometric technique similar to that used in \cite{aimeStresscontrolledShearCell2016}. This experimental setup uses the same laser beam employed for light scattering measurements, and described in deeper detail in section \ref{sec:rheooptics}.
The incoming beam illuminating the sample has an expanded radius of 2~cm and a converging spherical wavefront with radius of curvature $R_c=30$~cm. For small incident angles, untreated glass surfaces with a refractive index $n_{SiO_2}= 1.46$ have a reflectivity of $R_{air}=(n_{SiO_2}-1)^2/(n_{SiO_2}+1)^2=4\%$ when in contact with air, and $R_w=(n_{SiO_2}-n_w)^2/(n_{SiO_2}+n_w)^2=0.2\%$ when in contact with water-based samples with a higher refractive index, around $n_w=1.33$. 
We exploit the reflectivity of glass/air interfaces to build a simple interferometric setup, shown in Figure \ref{fig:parall}.

The beams reflected by the top and bottom interfaces are collected by a mirror and sent to a CMOS camera (CS165MU, from Thorlabs), as shown in Figure~\ref{fig:parall}. Before the camera, the two beams converge to a focal point. Adding a screen in that position, marked as a blue rectangle in Figure \ref{fig:parall}b, we visualize the beam positions in the transversal plane: their distance, $d_w$, reflects a misalignment $\alpha=d_w/(2R_c)$ between the two plates. Minimizing $d_w$ leads to a coarse alignment of the plates, with an accuracy set by the spot size, $w\sim 1$~mm, yielding a precision $\alpha_{coarse} \sim w/(2R_c)\sim 2.5\cdot 10^{-3}$~rad.
To improve the accuracy of the alignment, and quantitatively reconstruct the gap inhomogeneities, we remove the screen placed in the beam waist, and we let the two beams diverge for about 10~cm, and interfere on the camera detector, which records an interference pattern such as the one reported in Figure \ref{fig:parall}c. The observed set of fringes reveals spatial changes in the optical path difference between the two beams, arising from plate misalignment as well as from imperfections in their flatness. The distance between adjacent fringes in real space, $d_f$, corresponds to an accumulated optical path difference of one wavelength ($\lambda/n_w=400$~nm), therefore a gap mismatch $\delta_h=\lambda/(2n_w)=200$~nm.
The CMOS detector has 1440x1080 pixels, and a pixel size $l_p=3.45~\mu$m. The imaging system has a magnification $M_f=0.2$, allowing us to image a total area of about $25\times 19~mm^2$, with a resolution of $17.25~\mu$m. The corresponding range of detectable misalignment goes from a maximum of $\alpha_{Max}=M\delta_h/2l_p\approx 6\cdot 10^{-3}$~rad, about twice as large as the accuracy of coarse alignment, to a nominal minimum of $\alpha_{min}=M\delta_h/L_f\approx 10^{-5}$~rad, where $L_f\sim 3.5$~mm is the lateral size of the detector.
This overestimates the typical accuracy attained in experiments, where the flatness of the plate surfaces limits the alignment accuracy to about $\alpha_{typ}\approx 10^{-4}$~rad, as shown in Figure \ref{fig:parall}c. This alignment exceeds that offered by many commercially available rheometers, and will be considered as satisfactory in the following.

Once the two plates have been aligned at the target gap position, we raise the upper plate, we load $0.3-0.5~cm^3$ of the sample on the bottom plate of the shear cell, and we lower again the upper plate to the target position, to obtain a gap, $h$ typically ranging between $150~\mu$m and 1.5~mm, and a cross-sectional area, $S$, typically ranging between 1 and $10~cm^2$.
To convert raw force and displacement data provided by the air bearing stages, we need to accurately measure both $h$ and $S$. For both measurements, we exploit the optical microscope coupled to the rheo-optical setup, to take snapshots of the sample while the upper plate is lowered to the target position. For each image, taken when the vertical platform position is $\tilde{z}$, we track the sample edge and compute its cross-sectional area using ImageJ \cite{schindelinFijiOpensourcePlatform2012}. This procedure allows us to measure $S$ at the target position, with an uncertainty of about 0.05~cm$^2$, mostly due to the projection of the sample-air meniscus on the image plane. The measurement of $h$ is based on the motorized vertical stage, but requires a calibration of the motor position corresponding to contact. 
Without a normal force sensor, we cannot implement the calibration procedure routinely performed in commercial rheometers. Instead, we analyze $S(\tilde{z})$ obtained from microscopy images, and fit it with $S(\tilde{z})=V/(\tilde{z}-\tilde{z}_0)$, with $V$ the sample volume, and $\tilde{z}_0$ the zero-gap motor position. From the fit, we extract $\tilde{z}_0$, and we use it to obtain a measurement of the gap height $h=\tilde{z}-\tilde{z}_0$ at the target position, with a precision of about 10~$\mu$m, corresponding to a relative uncertainty of below 5\% on the derived viscoelastic moduli.





\section{Rheology of a microgel soft glass}\label{sec:validation}

To validate our mechanical calibration and to ensure the quality of the measurements obtained using our shear cell, we measure a sample with well-defined rheological properties and we compare results obtained with our shear cell with those from an Anton Paar MCR 302 rheometer equipped with a 25~mm cone-plate geometry with cone angle 2$^\circ$. Analysis of the raw data from the shear cell is done using an open source python package~\cite{GitHubRepo}
As a model sample, we use a well-characterized polyelectrolyte microgel suspension \cite{pelletGlassJammingTransitions2016} at a high polymer concentration, $C=3~wt\%$, beyond jamming, such that the sample behaves as a simple yield stress fluid.

\begin{figure}[ht]
    \centering
    \includegraphics[width=\columnwidth]{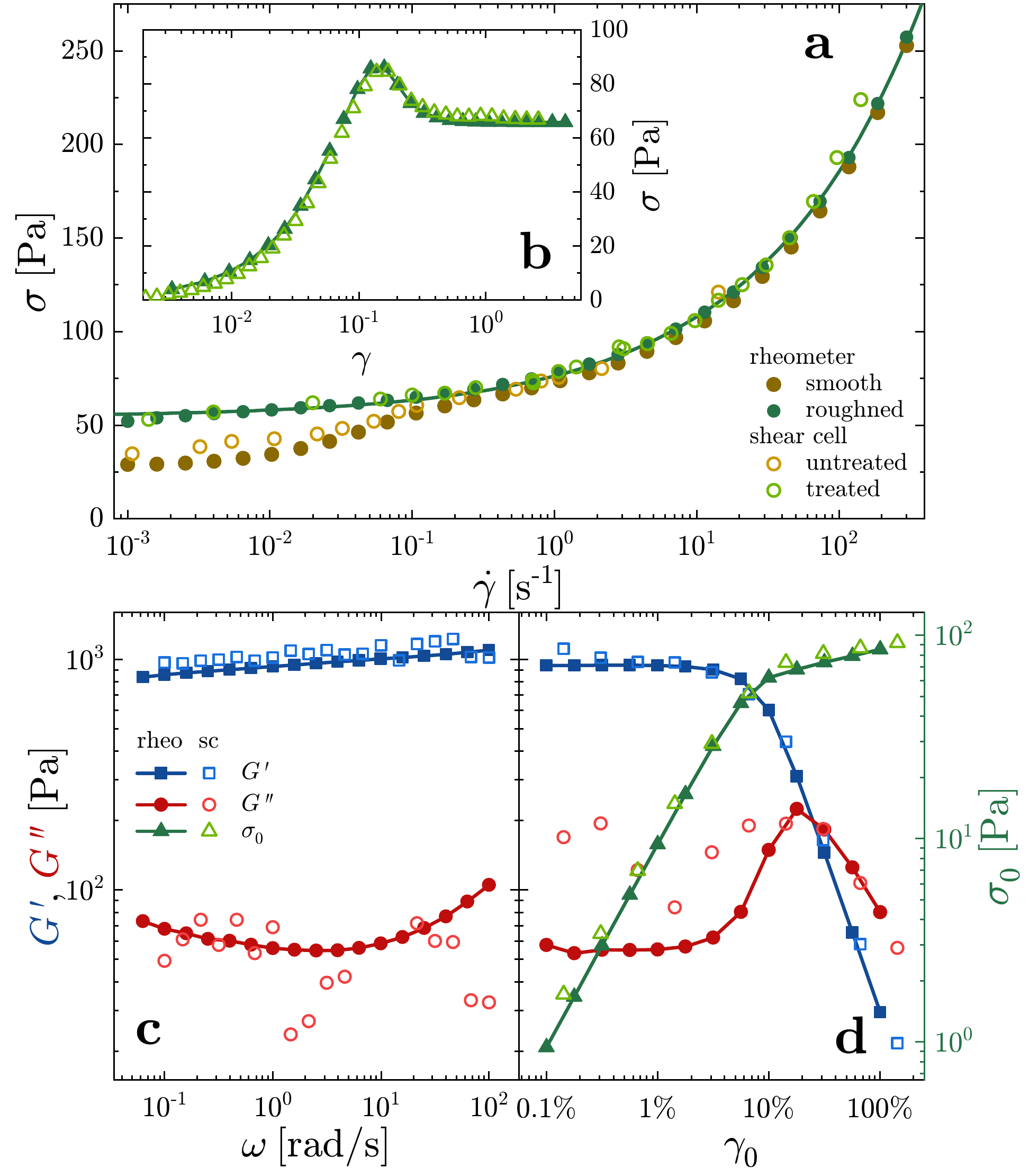}
    \caption[Rheonoslip]{\textbf{Rheology of a microgel soft glass} a) Full circles: flow curve of microgel soft glass measured with commercial rheometer with smooth (brown) and roughened cone-plate geometry (green). Green solid line: fit of green circles to the Herschel-Bulkley model. Open circles: flow curved measured with the shear cell with untreated (brown) and treated plates (green). b) Transient stress growth during shear startup measured for $\dot\gamma=0.17$~s$^{-1}$ with rheometer and shear cell (full and open symbols, respectively). c) Linear viscoelastic moduli $G^\prime$ (blue squares) and $G^{\prime\prime}$ (red circles) measured by strain-contolled oscillations at amplitude $\gamma_0=0.5\%$. d) Blue squares and red circles, left axis: First-harmonic viscoelastic moduli measured at $\omega=1$~rad/s. Green triangles, right axis: stress amplitude}
    \label{fig:Rheo}
\end{figure}

As a first characterization, we measure the steady-state flow curve, by imposing shear rates ranging between $10^{-3}~s^{-1}$ and $100~s^{-1}$, and measuring the steady-state stress value. At high shear rates, $\dot\gamma>1~s^{-1}$, we find that the stress measured with both instruments is well-described by the empirical Herschel-Bulkley (HB) model, $\sigma=\sigma_0+K\dot{\gamma}^n$, with $\sigma_0=50$~Pa, $n=0.4$, and $K=22~Pa\cdot s^n$, in good agreement with values reported in the literature for the same sample \cite{didioTransientDynamicsSoft2022}. At lower rates, we find that the measured stress is lower than the HB prediction, and slightly differs between the two instruments, suggesting the presence of wall slip~\cite{cloitreReviewWallSlip2017}. 
To test this hypothesis, we repeated the same experiment in the Anton Paar rheometer, using sandblasted geometries with a typical roughness of about $5 \mu m$. We obtain a flow curve, $\sigma(\dot\gamma)$, that confirms the behavior found at large $\dot\gamma$, and that approaches the HB plateau at low $\dot\gamma$, confirming that wall slip was the cause of the observed deviation, and that it was effectively suppressed in the roughened geometry, as shown by full green symbols in Figure \ref{fig:Rheo}a.
Mitigating wall slip in the shear cell is particularly challenging because the use of rough surfaces is incompatible with coupling the system to a light scattering setup. To overcome this, we applied a chemical surface modification to the plates, by immersing them in a 3\% water solution of polyethylenimine (PEI) 
for three days~\cite{christelStickslipControlCarbopol2012,chaub2024}. Repeating the flow curve measurement in the shear cell with treated plates, we obtain results in excellent agreement with the ones from the commercial rheometer, demonstrating that surface treatment effectively reduced wall slip, as shown by open green symbols in Figure \ref{fig:Rheo}a.

The good agreement of steady state stresses demonstrates that our shear cell is able to provide quantitatively accurate rheology results. To further benchmark its performance, we compare the time-dependent stress evolution measured by both instruments for one representative value of $\dot\gamma=0.17~s^{-1}$. The result, shown in Figure \ref{fig:Rheo}b, shows that the good agreement found at steady state extends to the transient regime as well, where our shear cell is able to capture the initial elastic response, characterized by an elastic modulus $G_0=1$~kPa, as well as the stress overshoot at yielding. By analyzing the time fluctuation of the stress signal at steady state, we extract a standard deviation $\sigma_\sigma=0.5$~Pa, estimating the uncertainty on individual stress points. This value corresponds to $\sigma_{noise}$ as derived in Sec.~\ref{sec:viscous}: it is about 10 times larger than that obtained with the commercial rheometer, $\sigma_{\sigma,rh}=45$~mPa, but good enough for most of the rheological tests that will be shown in the following.

To better assess the impact of stress noise on the outcome of rheology experiments, we perform oscillatory rheology at a fixed frequency, $\omega=1$~rad/s, and strain amplitude $\gamma_0$ varying between 0.1\% and 100\%. We measure raw force and displacement data on 20 oscillation periods, with an acquisition rate of 40 points per second. We process the raw data $\{F(t), x(t)\}$ by discarding the first oscillation period, computing the Fourier transform on an integer number, $N=16$, of periods to obtain $\{\hat{F}(\nu), \hat{x}(\nu)\}$, and restricting the analysis to the first-harmonic complex amplitudes, $\{\hat{F}_0, \hat{x}_0\}$, at $\nu=\omega$. 
From this, we compute stress and strain using the measured surface area, $S= 4~cm^2$, and sample gap, $h=700~\mu m$, and we extract the stress amplitude, $\sigma_0=|\hat{F}_0|/S$, as well as the complex viscoelastic modulus, $G^*=\hat{F}_0h/\hat{x}_0S$, whose real and imaginary part are $G^\prime$ and $G^{\prime\prime}$, respectively. 
The result shows the typical response of a yield stress fluid: at small strain amplitudes we find a solid-like behavior with a strain-independent $G'=G_0=1$~kPa, a noisier $G^{\prime\prime}$ about one decade smaller, and $\sigma_0\approx G_0\gamma_0$, as shown in Figure \ref{fig:Rheo}d. At larger $\gamma_0$, we observe yielding, with $G^\prime$ ultimately decreasing as $\gamma_0^{-1.3}$ and $G^{\prime\prime}$ becoming the dominant contribution, decreasing as $\gamma_0^{-0.6}$.
Comparing this result with that obtained in the commercial rheometer, we find good agreement apart from the value of $G^{\prime\prime}$ measured in the linear regime, for which shear cell data is noisy and about twice larger than that measured with the rheometer.
We find a similar trend on a wide range of $\omega$. Measuring the phase delay of the mechanical response, $\delta=\arctan(G^{\prime\prime}/G^{\prime})$, is easier for measurements at low frequency, $\omega<1$~rad/s, for which we obtain a cleaner measurement of $G^{\prime\prime}$ even in the linear regime, whereas $G^\prime$ is in good agreement with benchmark measurements up to the largest frequencies measured, $\omega_{Max}=100$~rad/s, as shown in Figure \ref{fig:Rheo}c.


Once validated that our device can accurately measure both transient and oscillatory shear rheology, we use it to perform superposition rheology experiments. To this end, we move one motor at a constant speed while the second motor performs small-amplitude oscillations with $\gamma_0=0.5\%$, to measure linear viscoelastic moduli during steady-state flow.
We perform this experiment with the two motors moving along perpendicular axes, to obtain the so-called Orthogonal Superposition Rheology (OSR) \cite{mewisMechanicalSpectroscopyColloidal1978,zeegersSensitiveDynamicViscometer1995,vermantOrthogonalSuperpositionMeasurements1997,vermantOrthogonalParallelSuperposition1998}.

\begin{figure}[ht]
    \centering
    \includegraphics[width=\columnwidth]{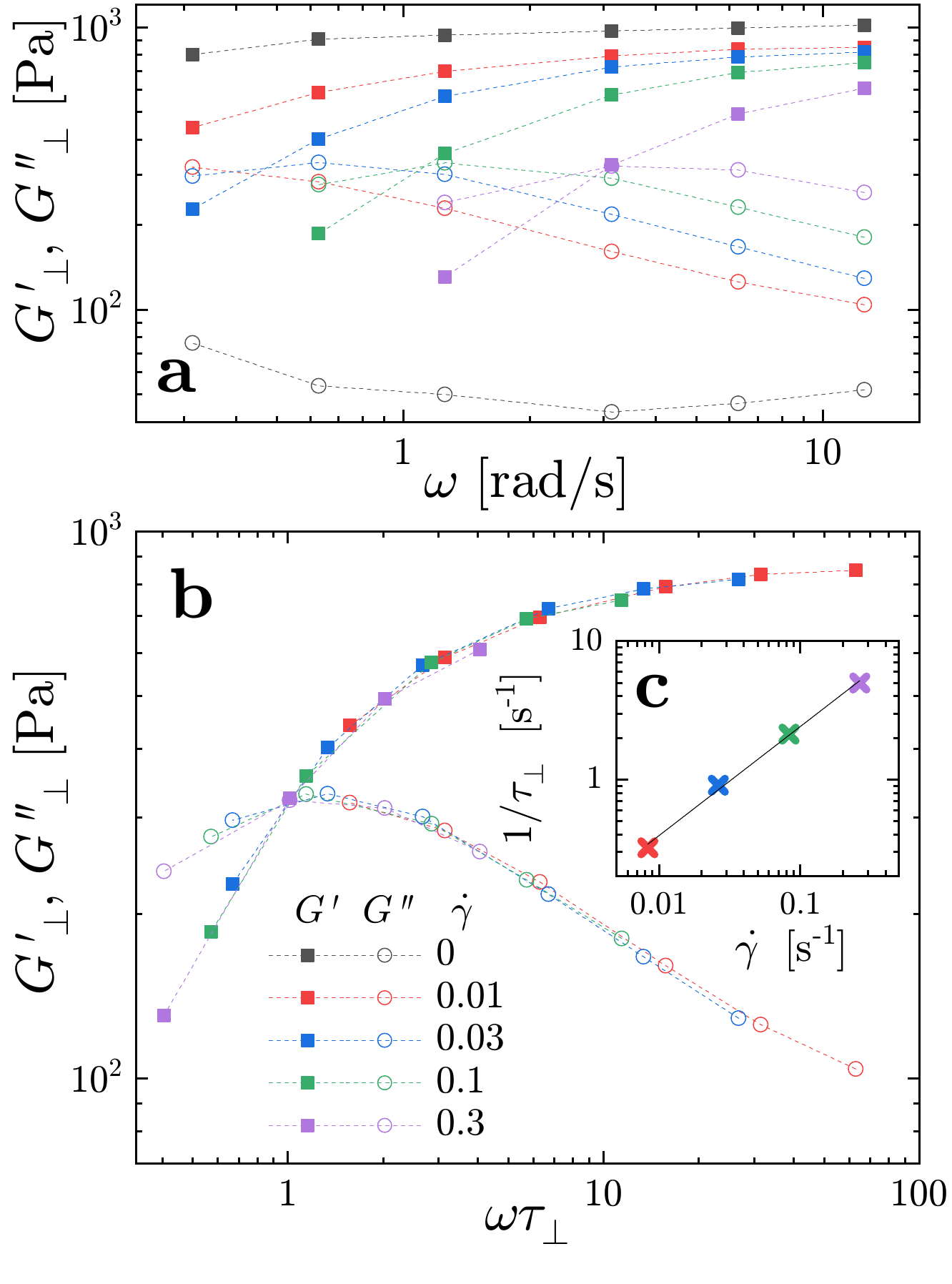}
    \caption{\textbf{Orthogonal Superposition Rheometry} a) Viscoelastic moduli, $G_\perp^\prime$ (full squares) and $G_\perp^{\prime\prime}$ (open circles) measured with the A-132 stage oriented along the $\hat{y}$ axis during steady-state shear imposed by the A-131 stage along $\hat{x}$, for different shear rates $\dot\gamma$, increasing from blue to red shades, as specified in the legend. b) Same data as in panel a, rescaled horizontally by a factor $\tau_\perp$ to collapse them on a mastercurve characterized by the crossover of $G_\perp^\prime$ and $G_\perp^{\prime\prime}$ at $\omega\tau_\perp=1$. Data for $\dot\gamma=0$ is not shown. c) characteristic relaxation frequency, $1/\tau_\perp$, as a function of the imposed rate.}
    \label{fig:osr}
\end{figure}

Even at the smallest shear rates measured, we find that the linear viscoelastic spectrum differs significantly from the one measured with the sample at rest. In particular, we find that the difference is largest at low frequencies, where $G^\prime$ and $G^{\prime\prime}$ approach a crossover, at frequency $\omega_X$, that enters the measured range of frequencies for $\dot\gamma\geq 0.03$~s$^{-1}$, and increases with increasing $\dot\gamma$, as shown in Figure \ref{fig:osr}a. This indicates the emergence of a shear-induced characteristic relaxation time, $\tau_\perp=1/\omega_X$, that was absent in the viscoelastic spectrum measured at rest. 
We find that the entire shear-rate dependence observed in Figure~\ref{fig:osr}a can be described in terms of this relaxation time, as demonstrated by the nice collapse of all viscoelastic spectra on a mastercurve when plotted against the rescaled frequency $\omega\tau_\perp$. This result is in good agreement with what previously found in colloidal glasses, where it was interpreted as the signature of microscopic shear-induced rearrangements involving particles exchanging their nearest neighbors and dissipating elastic energy, a mechanism called convective cage release~\cite{jacobConvectiveCageRelease2015}. 

By contrast, repeating the same experiment with the two motors moving along parallel axes, we perform Parallel Superposition Rheology (PSR) experiments. These experiments are technically simpler, and also accessible to most torsional rheometers, but conceptually harder to interpret, due to nontrivial coupling between the two imposed deformation profiles \cite{dhontSuperpositionRheology2001}. Accordingly, our PSR results differ from OSR ones, as the viscoelastic moduli measured under different shear rates no longer collapse on a mastercurve upon horizontal rescaling, as shown in Appendix~\ref{secA:PSR}. This is one example highlighting the value of 2D shear experiments allowed by our two-axes shear cell. 

\section{Rheo-optics}\label{sec:rheooptics}

To further illustrate the potential of our shear cell, we couple it to light scattering and microscopy as described in Section \ref{sec:setup}, and we exploit this rheo-optical platform to study in deeper detail the microscopic origin of the observed rheological behavior.
To this end, we illuminate the sample with the expanded beam of a solid-state laser (Genesis MX532-1000 from Coherent), with a wavelength $\lambda=532$~nm and a maximum output power of 1W, and we dope the microgel suspension with a tiny amount of polystyrene tracer particles (Fluoro-Max, from Thermo Fisher), labeled with a Rhodamine B fluorophore with an excitation peak wavelength, $\lambda_{ex}=541$~nm, close to the wavelength of the laser used for light scattering. This allows us to perform both Particle Tracking Velocimetry (PTV) and Photon Correlation Imaging (PCI).

\subsection{\label{sec:PTV}Particle Tracking Velocimetry}

\begin{figure}[ht]
    \centering
    \includegraphics[width=\columnwidth]{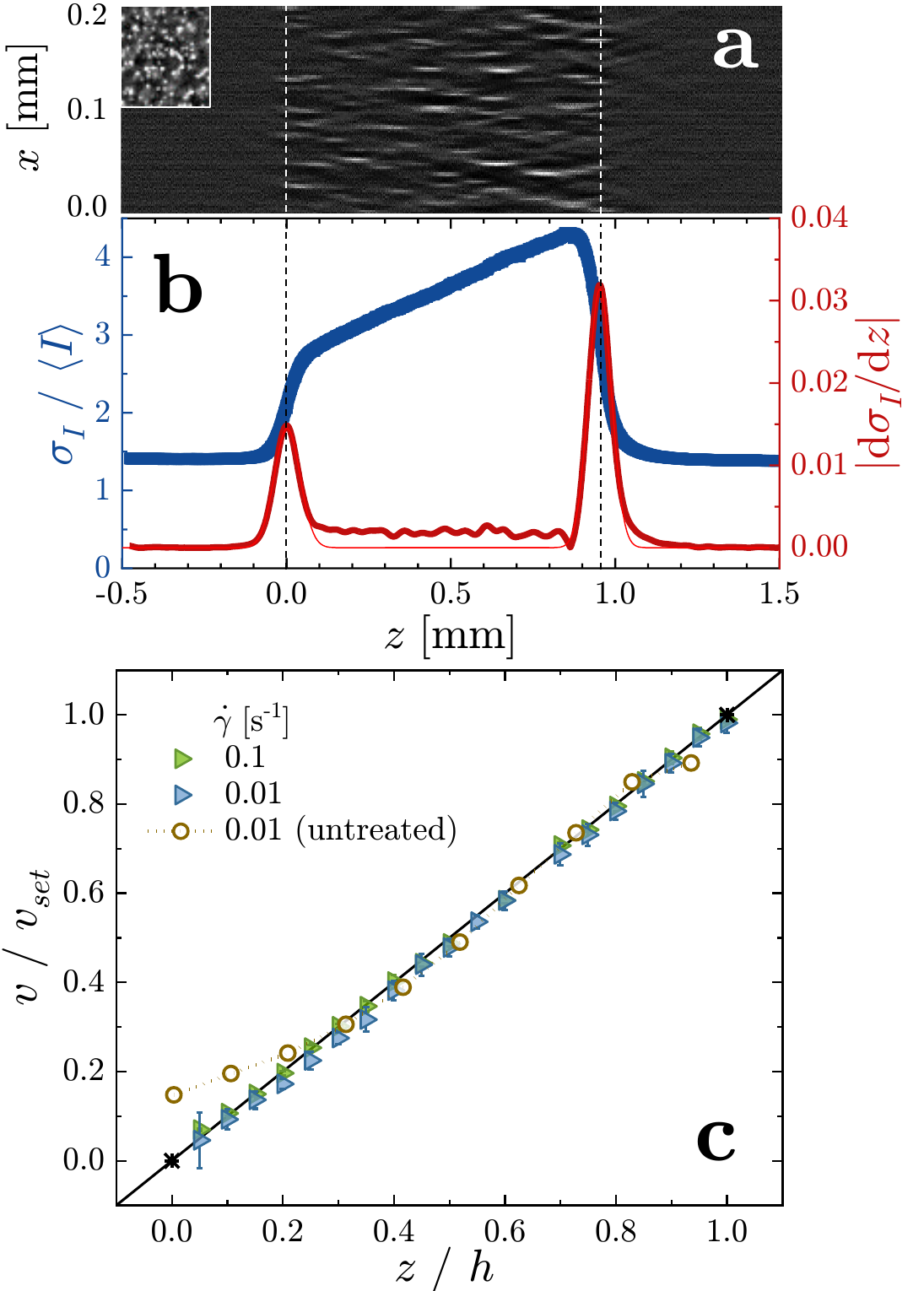}
    \caption[Gapfluoresent]{\textbf{Rheo-PTV} a) example of fluorescence microscopy image stack projected on the $xz$ plane. White dashed vertical lines mark the sample edges, beyond which no fluorescent particle is detected. The inset reports a detail of the microscopy image in the $xy$ plane. b) blue, left axis: standard deviation of grayscale intensity detected in the $xy$ plane, normalized by the average intensity, $\sigma_I/\langle I \rangle$. Red, right axis: $z$ derivative of $\sigma_I$ (dark red line) fitted by two Gaussian peaks (light red line) with peak position corresponding to sample edges and FWHM $\Delta_z=80~\mu$m, quantifying the depth of focus of the microscope. c) Full triangles: drift speed as a function of $z$, normalized by the speed of the top plate, $v/v_{set}$, for two values of $v_{set}$, corresponding to shear rates $\dot\gamma=0.1$ and 0.01~s$^{-1}$, as specified in the legend. Open brown circles: flow profile measured with untreated surfaces, giving rise to wall slip. Black asterisks: imposed boundary conditions. Solid line: expected affine deformation field.} 
   \label{fig:gapfluorescent}
\end{figure}

For PTV experiments, we equip the microscope with a notch filter (NF533-17, from Thorlabs), to filter out the laser wavelength, such that only fluorescent light emitted by the particles is imaged by the microscope camera.
In this optical configuration, particles are well visible, with a Point Spread Function of about 2.8$\mu$m in lateral size at the largest magnification, $3.5\times$. 
We then use the motorized focusing of the AxioZoom to move the object plane across the sample, taking pictures in steps of $\delta z_m=1~\mu$m, corresponding to $\delta z_s=\delta z_m/n=0.75~\mu$m in the sample, $n\approx 1.33$ being the refractive index. As the object plane is moved out of the sample gap, the image of the tracers becomes blurred and fades out. This blurring, well visible in the image stack projected on the $xz$ plane, as shown in Figure~\ref{fig:gapfluorescent}a, can be quantified by computing the spatial variance of the grayscale distribution, $\sigma_I^2$, normalized by the averaged intensity, $\langle I\rangle^2$, as a function of $z$ position of the imaged plane. This allows us to identify the sample edges, which we identify with the inflection points of $\sigma_I/\langle I\rangle$, as shown in Figure~\ref{fig:gapfluorescent}b.
With the sample at rest, we record no significant motion of the tracer particles, as expected due to the highly jammed state of the sample along with the low resolution of the imaging system.
By contrast, under shear, we detect a uniform drift of the particles, which depends on the position of the plane being imaged. We analyze this drift motion by tracking individual particles using the trackpy package \cite{allan_trackpy:_2015}, and averaging the motion of $\approx 10^3$ particles per frame to obtain the flow speed at one given position, $z$, across the gap. We then repeat this measurement for different planes to reconstruct the flow profile, $v(z)$. 
To optimize the number of independent planes that can be measured, we estimate the depth of focus of the microscope by fitting the peaks of $d\sigma_I/dz$ to Gaussian peaks, as shown in Figure~\ref{fig:gapfluorescent}b. The result yields a full width at half maximum (FWHM) of $\Delta_z=80~\mu$m. 
This value is larger than the $z$ resolution of rheo-microscopy setups employing objectives with higher numerical aperture~\cite{villaQuantitativeRheomicroscopySoft2022,ederaDeformationProfilesMicroscopic2021}, which could not be used in this setup due to their much smaller working distance.
To increase the $z$ resolution, we refine our PTV analysis by tracking only 10\% of the observable particles, those characterized by high brightness and small size, thereby discarding particles out of focus. We estimate that this procedure improves our $z$ resolution by a factor $\approx 2$: therefore, we resolve our PTV analysis on 20 planes, with a spacing of about 35~$\mu$m. 
For all measured shear rates, we find that the velocity profile is linear and that $v$ extrapolates to 0 at the fixed plate position and to the imposed plate speed at the moving plate position. By contrast, repeating this analysis with untreated shear cell plates, for which we measure the anomalous flow curves reported as brown shades in Figure~\ref{fig:Rheo}a, we obtain significant deviation from the expected affine behavior, as shown by open circles in Figure~\ref{fig:gapfluorescent}c. This results provides direct evidence of the wall slip mechanism evoked to explain the observed anomalous flow curves, while confirming the absence of wall slip and shear banding when using surface-treated plates.

\subsection{\label{sec:PCI}Photon Correlation Imaging}

\begin{figure}[ht]
    \centering
    \includegraphics[width=\columnwidth]{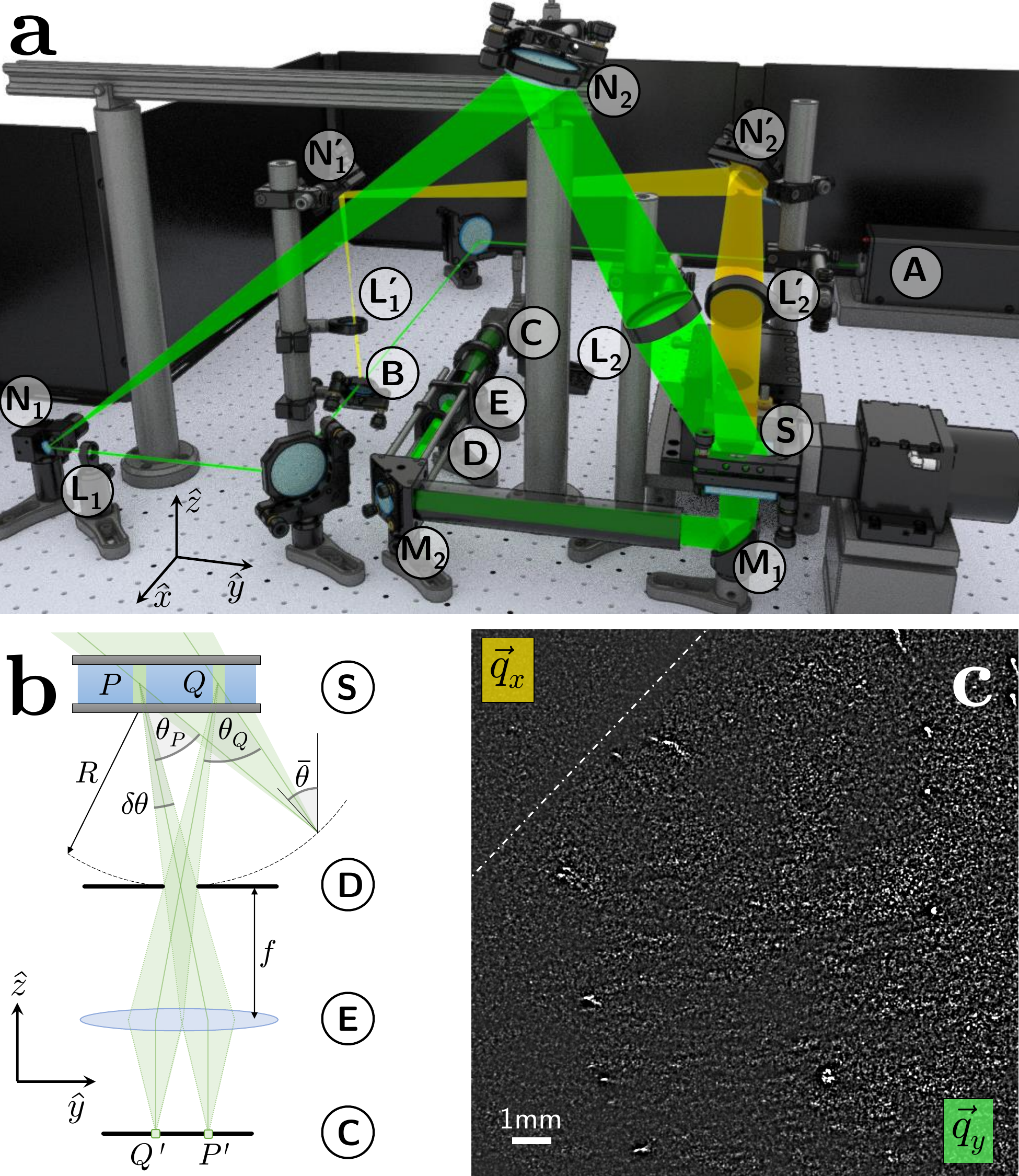}
    \caption[sketchDLS]{\textbf{Rheo-PCI} a) Sketch of the optical setup used for PCI experiments under shear: ($A$) Laser; ($B$) Beam splitter; ($L_{1,2}$) beam-shaping lenses for $yz$ scattering plane; ($L_{1,2}^\prime$) beam-shaping lenses for $xz$ scattering plane; ($N_{1,2}$, $N_{1,2}^\prime$) mirrors setting the scattering angle; ($S$) Sample in the shear cell; ($M_{1,2}$) mirrors collecting the scattered light and sending it to the PCI camera ($C$) through a diaphragm ($D$) and a lens ($E$). b) Sketch of the PCI geometry, for the scattering plane $yz$ (green incoming beam in panel a). c) Example of the speckle pattern detected by the CMOS camera. The white dashed line highlights the border between the regions illuminated by the two incoming beams, which is barely visible in the image as the two beams are nearly overlapping.}
    \label{fig:sketchDLS}
\end{figure}

We complement these observations with the measurement of microscopic dynamics at length scales comparable with, or even smaller than, the size of individual particles, using the PCI setup shown in Figure~\ref{fig:sketchDLS}. 
For PCI experiments, we illuminate the sample with an expanded laser beam, with an incident angle relative to the vertical $\hat{z}$ axis, $\bar\theta$, adjustable between $20^\circ$ and $60^\circ$ thanks to a galvanometric scanner mirror (6215H with servo controller DC900, from Cambridge Technology), controlled by a reference voltage signal from a DAQ interface module (USB-6001, from National Instruments). 
The scattered light is then collected from below by a mirror, filtered by a diaphragm aperture to tune the numerical aperture of the collection system in the range $NA\in [0.002,0.02]$, and sent to a CMOS sensor (Basler aca2040-180km) which records an image of the sample with magnification $M_{PCI}=0.4$.
The detector has $2048\times 2048$ pixels, with pixel size $l_p=5.5~\mu$m. The low magnification enables imaging a large field of view, $L_{PCI}=2048 l_p/M\approx 2.8$~cm, typically including the whole sample being sheared. Under the assumption of Gaussian illumination and collection optics, the imaging system has a Point Spread Function (PSF) of lateral size $\Sigma\sim\lambda M/NA$, larger than the pixel size in the whole range of accessible $NA$. As a result, the camera records a speckle pattern, each speckle representing the intensity scattered by particles in a tiny scattering volume, of lateral size $\tilde\Sigma=\Sigma/M$ and spanning the entire sample thickness. The temporal fluctuation of the speckle intensity reflects relative motion projected on the scattering vector, $\vec{q}$, which has a magnitude $q=2k\sin(\bar\theta/2)$, with $k=2\pi n/\lambda\approx 15.7 \mu m^{-1}$ the laser wavevector in the scattering medium and $\bar\theta$ the scattering angle. Orienting $\vec{q}$ perpendicular to the shear velocity, we obtain selective sensitivity to non-affine microscopic dynamics, over length scales $2\pi/q$ of a few hundred nanometers, which we can resolve in space with a resolution set by $\tilde\Sigma$ over the entire sample area.

In this geometry, ensuring that the scattering angle is the same for all sample locations is far from trivial. Using the plane-wave illumination typical of most light scattering setups sets a well-defined wavevector $\vec{k}_{in}$ for the incoming laser beam. However, the diaphragm in the collection optics filters light scattered at an angle which slightly depends on the sample location, resulting in spatial variations of the scattering vector, $\vec{q}=\vec{k}_{out}-\vec{k}_{in}$. For a sample with a typical radius of order $r\sim 1$~cm, the spread in scattering angle is about $\delta\theta = r/R\sim 30$~mrad, $R=30$~cm being the optical path between the sample and the diaphragm. Albeit small, the resulting spread in scattering vector, $\delta q \approx k\cdot \delta\theta\sim 0.5~\mu$m$^{-1}$, has a strong impact on the light scattering signal, which is highly sensitive on spurious components of $\vec{q}$ along the shear direction \cite{aimeProbingShearinducedRearrangements2019,pommellaCouplingSpaceResolvedDynamic2019}.
To minimize this spread, we illuminate the sample with a converging spherical wavefront, with radius of curvature $R_c=R$. In this geometry, the incoming wavevectors illuminating different sample regions, $\vec{k}_{in}$, are oriented along a direction that slightly differs between different sample location, in a way that matches the difference in the scattered wavevectors, $\vec{k}_{out}$, collected by the lens through the diaphragm aperture, minimizing the spatial variations of the scattering vector, $\vec{q}=\vec{k}_{out}-\vec{k}_{in}$, as shown in Figure \ref{fig:sketchDLS}b.

To simultaneously detect dynamics along the velocity direction, which is precious for instance to detect the presence of shear banding or wall slip, we split the incoming laser beam and add a second scattering vector, of similar magnitude, but oriented on a scattering plane perpendicular to the first one, as schematically shown in yellow in Fig. \ref{fig:sketchDLS}a. 
To avoid illuminating the same scatterers with two laser beams at the same time, two strategies are possible: the first one, implemented in \cite{aimeDynamicSpeckleHolography2021}, is to strobe the light, alternating illumination from the two directions, and to separate even and odd frames when postprocessing the acquired videos. Here, to maximize the acquisition rate, we adopt a different strategy: using slits (not shown for clarity in Fig. \ref{fig:sketchDLS}), we block half of each illuminating beam, thereby splitting the field of view of the camera in two parts, each one illuminated by one incoming laser beam. 
A representative speckle field resulting from the juxtaposition of the two scattered beams is shown in Fig.~\ref{fig:sketchDLS}c. The speckle pattern in the upper right corner of the picture is generated by particles illuminated by the yellow beam in Fig.~\ref{fig:sketchDLS}a, and has a scattering vector $\vec{q}_x$, mostly oriented along the velocity direction. By contrast, the speckle pattern in the remaining bottom right part of the picture comes from illumination in the $yz$ plane, and has a scattering vector $\vec{q}_y$, mostly oriented along the vorticity direction, and with no component along the velocity direction. The slits are carefully positioned such that the two speckle patterns do not overlap, leaving a thin darker separating region highlighted by the dash-dotted line in Fig.~\ref{fig:sketchDLS}c.
We then analyze the two regions independently to gain information on motion along the two directions. 
Finally, to improve sensitivity to dynamics along the sher gradient, we remove the notch filter from the AxioZoom objective, we decrease its numerical aperture, and we use it to image a speckle pattern similar to the one previously described, carrying information on light scattered at a wide angle, $\theta^\prime=\pi-\theta$, such that the scattering vector is mostly oriented along the shear gradient direction. The two cameras are then acquired simultaneously using a dual-camera frame grabber (Matrox Radient eV-CL), and processed following the data analysis protocol described in \cite{aimeDynamicSpeckleHolography2021}.

\begin{figure}[h]
\centering
\includegraphics[width=\columnwidth]{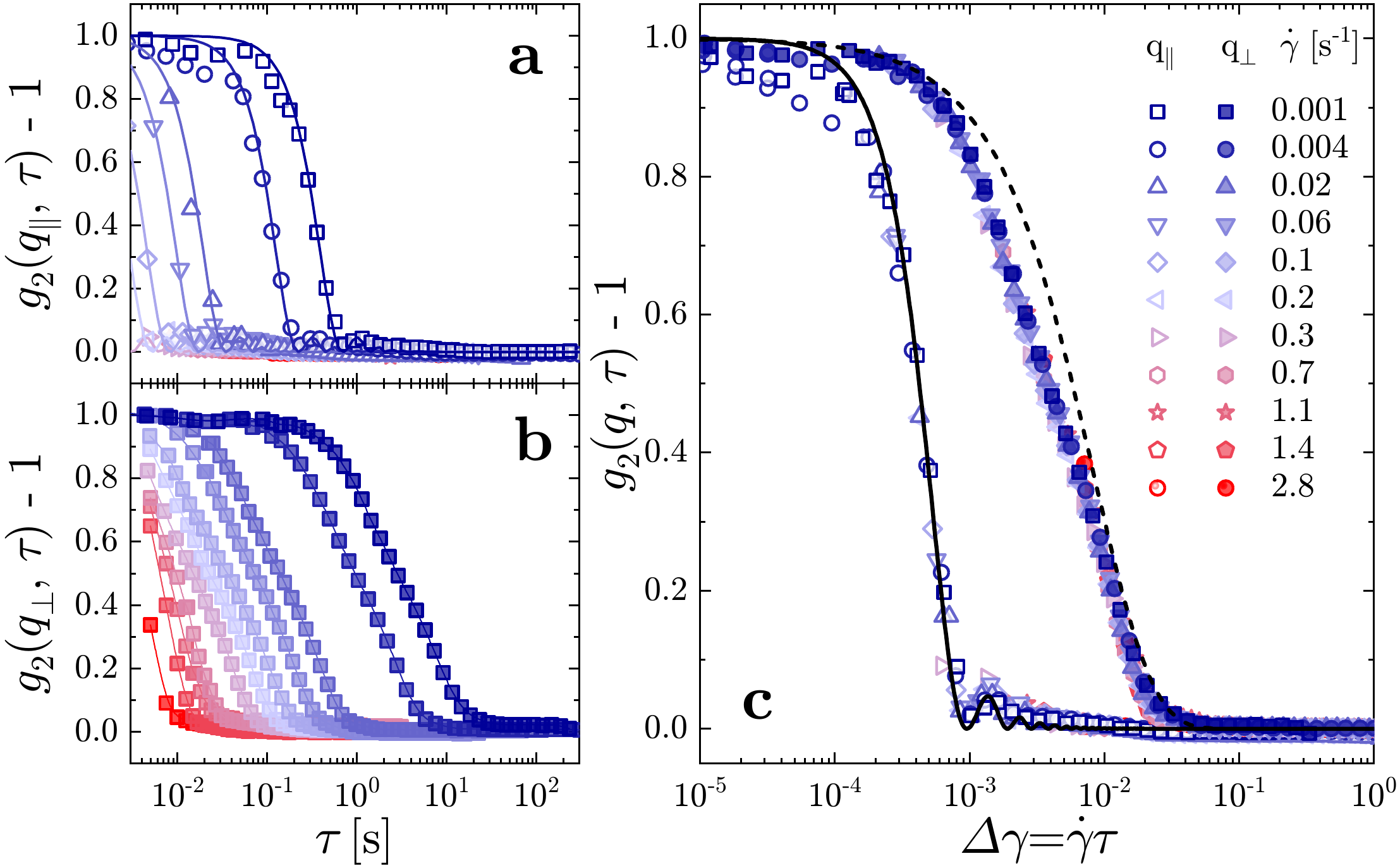}
\caption{\textbf{Microscopic dynamics under shear} a) Symbols: intensity correlation functions measured under steady shear at rate $\dot\gamma$, increasing from blue to red shades, for a scattering vector $q_\parallel=10~\mu$m$^{-1}$ mostly oriented along the flow direction. Lines: theoretical model based on the assumption of purely affine deformation. b) correlation functions measured for $q_\perp=10~\mu$m$^{-1}$ oriented perpendicular to the flow direction, and mostly along the vorticity direction. c) Symbols: same data as panels a and b, plotted agains the incremental strain $\Delta \gamma$. Black solid line: correlation function from purely affine dynamics. Dashed black line: parameter-free theoretical model based on shear-induced diffusion with diffucion coefficient $\tilde{D}=a^2\dot\gamma/\gamma_y$}\label{fig:corrfuncs}
\end{figure}

A detailed analysis of the scattering signal goes far beyond the scope of this paper. Here, we demonstrate its potential by probing affine and nonaffine dynamics under steady-state shear at various rates $\dot\gamma$, imposed along $\hat{x}$ using the A-131 stage. This analysis highlights the dynamics of the same tracer particles used for PTV experiments, whose scattering signal is way stronger than that of the microgel particles. For each $\dot\gamma$, we take sequences of up to $10^4$ images using a non-uniform time sampling algorithm designed to probe a large range of time scales at a moderate average acquisition rate~\cite{philippeEfficientSchemeSampling2016}. We then use a python package~\cite{DSHpy} to compute affine and nonaffine correlation functions by averaging time-resolved correlations~\cite{duriTimeresolvedcorrelationMeasurementsTemporally2005a} on speckles belonging to the $\vec{q}_x$ and $\vec{q}_y$ image regions, respectively, as shown in Figure~\ref{fig:sketchDLS}c.
Restricting the analysis to steady state flow, for $\gamma=\dot\gamma t>0.5$, we observe that dynamics are stationary, and we average them in time to obtain intensity correlation functions, $g_2-1$, shown in Figure~\ref{fig:corrfuncs}.
In the direction parallel to shear ($\vec{q}_x$), all correlation functions are dominated by the affine deformation: we find that $g_2(q_\parallel, \tau) - 1 = \textrm{sinc}^2(q_\parallel h \dot\gamma\tau/2)$, where $\textrm{sinc}(x)\equiv\sin(x)/x$, as expected for purely affine dynamics~\cite{aimeProbingShearinducedRearrangements2019}.
This is better visualized by rescaling $\tau$ using the imposed $\dot\gamma$, and plotting $g_2-1$ as a function of the incremental strain, $\Delta\gamma=\dot\gamma\tau$, as shown by the open symbols in Figure~\ref{fig:corrfuncs}c. 
In the perpendicular direction ($\vec{q}_y$), we measure a $\dot\gamma$-dependent correlation decay, indicating that motion at the microscale is not purely affine. We interpret this decay as the signature of shear-induced microstructural rearrangements responsible for stress dissipation during viscoplastic flow \cite{khabazParticleDynamicsPredicts2020}
We observe that correlation functions measured at different $\dot\gamma$ have all the same shape, and they perfectly collapse on a mastercurve when plotted against $\Delta\gamma$, as shown by the full symbols in Figure~\ref{fig:corrfuncs}c. This suggests that the dynamics measured at different shear rates differ only by the imposed timescale, $1/\dot\gamma$, analogous to the collapse of OSR viscoelastic moduli shown in Figure~\ref{fig:osr}b.
Building upon this analogy, we use our light scattering data to confirm the convective cage release mechanism proposed in the literature to explain similar OSR results \cite{jacobConvectiveCageRelease2015}. This model associates yielding to particles escaping the cage formed by their neighbors. We assume that these cage-escaping events take place with a characteristic frequency set by $\dot\gamma/\gamma_y$, the inverse of the time required to reach the yield strain, $\gamma_y\sim 15\%$, as defined by the stress overshoot shown in Figure~\ref{fig:Rheo}b. We further assume that these events entail particle motion over a fistance of the order of the particle diameter, $a=0.3~\mu$m. If shear-induced dynamics are diffusive, as suggested by previous experiments on soft colloidal glasses~\cite{aimeUnifiedStateDiagram2023,ederaYieldingMicroscopeMultiscale2024} and simulations~\cite{mohanLocalMobilityMicrostructure2013,khabazParticleDynamicsPredicts2020}, the correlation function should then be: $g_2(q_\perp, \tau)-1=\exp(-2\tilde{D}q_\perp^2\tau)=\exp(-2a^2q_\perp^2\Delta\gamma/\gamma_y)$, with $\tilde{D}=a^2\dot\gamma/\gamma_y$ the shear-induced diffusion coefficient.
We find that this naive assumption predicts an exponential decay with a decay strain $\gamma_R=\gamma_y/(2a^2q_\perp^2)=0.8\%$, which captures quite well, without adjustable parameters, the observed behavior, as shown by the dashed black line in Figure~\ref{fig:corrfuncs}c, confirming the microscopic model introduced to explain our OSR data, and demonstrating the potential of coupling superposition rheology and light scattering.

\section{Conclusion}\label{sec:conclusion}
In this work, we have developed a novel two-axes shear cell based on commercially-available air bearing stages, which we have calibrated and optimized to perform 2D shear rheology. Natively stress-controlled, our shear cell can operate in strain-controlled mode through a servo loop integrated into the dual-axis stage controller. It can apply or measure stresses ranging between about 1 and $10^4$~Pa, and strains up to $\sim 10$ strain units, with a frequency bandwidth set by several competing factors that we carefully characterized. This makes our shear cell well suited for a broad range of soft materials.
We validated the shear cell by measuring a microgel soft glass, finding good agreement with data from a commercial rheometer. 
Exploiting its two independent axes, we conducted orthogonal superposition rheology experiments, finding the signature of convective cage release events characterized by a shear-induced relaxation time $\tau \propto \dot\gamma^{-1}$.
To demonstrate the potential of our sher cell, we coupled it to an optical microscope and a photon correlation imaging (PCI) setup, to trace back the rheological signal to its microscopic origin. We used the microscope to perform particle tracking velocimetry under shear, and confirmed the presence of wall slip in samples exhibiting anomalous flow curves. 
In absence of slip, we used PCI to measure affine and nonaffine dynamics taking place at the scale of a single colloidal particle under steady-state flow. We found the signature of shear-induced diffusion, with a diffusion coefficient simply proportional to the shear rate, the proportionality coefficient being set by the particle size and the yield strain. 

Ongoing research in our group couples this shear cell to other techniques, including photochemistry, UV-visible spectroscopy, fluorescence microscopy to detect the stress-activated signal of mechanophores, and dynamic speckle holography.
We have also adapted simpler version of the shear cell, with one single axis, to perform uniaxial compression, three-point bending, and tensile tests for the study of porous materials and elastomers.

Overall, this shear cell provides a flexible platform to extend rheological measurements, coupling them to a variety of complementary techniques to unveil the microscopic origin of complex rheological behavior. In particular, the combination of rheology, microscopy and photon correlation imaging presented here holds a great potential for investigating the elusive onset of nonlinear rheology, with applications to memory effects, fatigue, yielding and failure of soft materials.

\backmatter
\bmhead{Author contributions} SA designed and built the setup, SA and CM performed experiments and analyzed the data. SA wrote the paper.
\bmhead{Acknowledgements} We are thankful to Michel Cloitre for insightful discussions, for providing the colloidal sample, and for help with the interpretation of rheology data, especially in presence of wall slip. We thank Jean-Marc Suau from Arkema for microgel particle synthesis. We thank Arnaud Chaub for help with surface treatment and with Rheo-PIV experiments. We thank Ludovic Olanier and Jean-Claude Mancer for help with the realization of mechanical components. We thank Gonzalo Sanchez Vera for help moving the rheo-optical setup when the entire lab was moved to the new building. SA thanks Francois Tournilhac and Shmuel Rubinstein for insightful discussions that led to setup improvements, and Matt Reck from the air bearing department at Phisik Instrumente, for invaluable help troubleshooting the air bearing stage, and for agreeing to develop the prototype A-132 stage upon our request. 
\bmhead{Funding} This work was supported by European Union's Horizon 2020 research and innovation programme under the Marie Sklodowska-Curie grant agreement No 955605 (YIELDGAP), and by French DIM programme 2020-32 (RESPORE)
\bmhead{Conflict of interest} The authors have no competing interests to declare that are relevant to the content of this article.
\bmhead{Data availability} The datasets generated during and/or analysed during the current study are available from the corresponding author on reasonable request. The code used to analyze the data is available on GitHub \cite{GitHubRepo}.

\begin{appendices}

\section{Current measurement}\label{secA:controller}

The servo controller adjusts the current flowing in the voice coil, $I$, using a feedback loop updated at a frequency of 20~kHz. The current set point and actual output from the feedback loop are measured in units of the motor peak current ($I_{Max}=10$~A for the motors used here) every loop cycle, and stored as floating-point internal variables in the controller memory. Those variables, named \verb|SP0:axes[N].command| and \verb|SP0:axes[N].iq| respectively, with \verb|N=0,1| denoting the two motors, can be accessed by custom-made scripts running on dedicated buffers of the controller firmware, as exemplified hereafter for the first motor (A-131):

\begin{verbatim}
REAL sp_add_iq, sp_add_cmd
add_iq  = GETSPA(0, "axes[0].iq")
add_cmd = GETSPA(0, "axes[0].command")
WHILE 1
    IQ_AMPS(0)   = 10 * GETSP(0, add_iq)
    ICMD_AMPS(0) = 10 * GETSP(0, add_cmd)
END
\end{verbatim}

Other three variables stored in the controlled memory are potentially useful for current measurement: 
\begin{itemize}
    \item \verb|RMS|, measuring RMS current averaged over 100~ms time windows, in units of percent of $I_{Max}$, and stored as a floating-point number ranging from 0 to 100
    \item \verb|DOUT|, measuring current output as a 16-bit integer variable, mapping the interval $[I_{Max},-I_{Max}]$ to $[-2^{15},2^{15}-1]$, positive integers reporting negative currents as per other variables definition
    \item \verb|DCOM|, only updated when the motor is driven in open-loop (force-controlled) mode, reporting the current set-point in units of $I_{Max}$, as a floating-point variable
\end{itemize}

Based on this assessment, we measure $I$ for each motor once per controller cycle, and store in variables, \verb|I0| and \verb|I1|, defined as \verb|0.1*DCOM| in open loop mode and \verb|-10/32767*DOUT| in closed loop mode. \verb|I0| and \verb|I1| are then saved at a lower frequency, depending on the experimental time scale. To smooth out fluctuations and improve the accuracy of sparse sampling, we implement a controller routine averaging current readings over the last $N$ controller cycles. This is done through the following single-line routine (here indented to improve readability) that is run once per controller cycle:

\begin{verbatim}
WHILE 1;
    IF MST0.#OPEN = 1; 
        x = DCOM0 * 0.1; 
    ELSE; 
        x = -10.0 * DOUT0 / 32767; 
    END; 
    y = DSHIFT(I0buf, x, N-1); 
    I0 = I0 + (x - y) / N;
END
\end{verbatim}

Here, larger $N$ produce smoother output variables, but effectively add a time delay between the instantaneous position and the time-averaged current. To compromise between signal-to-noise ratio and timing accuracy, we choose $N=4$ throughout this paper.

\section{Parallel Superposition Rheology}\label{secA:PSR}

\begin{figure}[ht]
    \centering
    \includegraphics[width=\columnwidth]{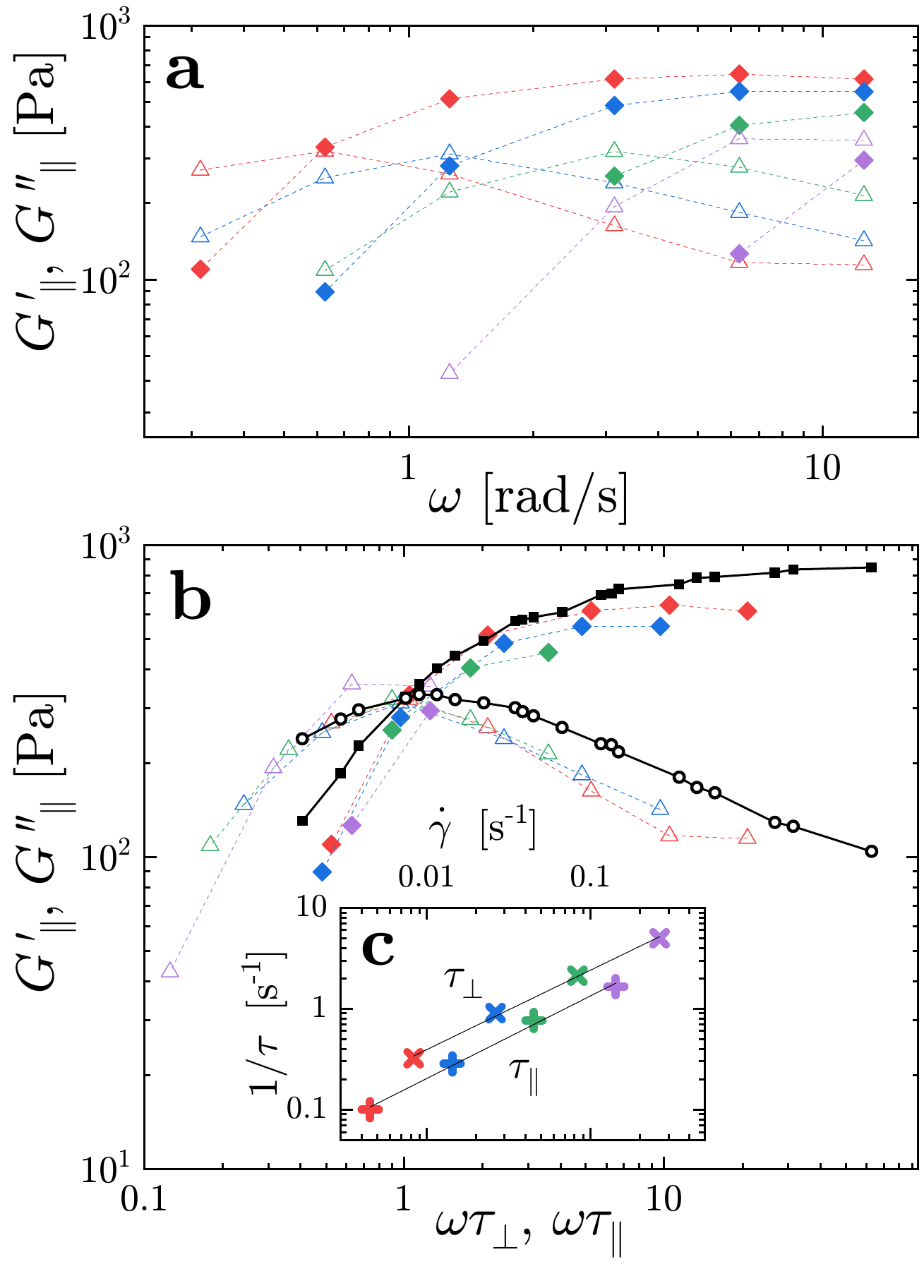}
    \caption{\textbf{Parallel Superposition Rheometry} a) Viscoelastic moduli, $G_\parallel^\prime$ (full diamonds) and $G_\parallel^{\prime\prime}$ (open triangles) measured with the A-132 stage oriented along $-\hat{x}$ during steady-state shear imposed by the A-131 stage along $\hat{x}$, for different shear rates $\dot\gamma$, increasing from blue to red shades. b) Colored symbols: same data as in panel a, rescaled horizontally by a factor $\tau_\parallel$ so that the crossover between $G_\parallel^\prime$ and $G_\parallel^{\prime\prime}$ takes place at $\omega\tau_\parallel=1$. Black open symbols: mastercurve obtained by the collapse of orthogonal superposition moduli, reported in figure \ref{fig:osr}. c) characteristic relaxation frequencies, $1/\tau_\perp$ ($\times$) and $1/\tau_\parallel$ ($+$), as a function of the imposed rate.}
    \label{fig:psr}
\end{figure}

We report in Figure~\ref{fig:psr} the result of Parallel Superposition Rheology (PSR) experiments, performed repeating the experimental protocol of Orthogonal Superposition Rheometry (OSR) described in Section~\ref{sec:validation}, with the two motors moving along parallel axes. We find that our PSR results differ from OSR ones, as the viscoelastic moduli measured under different shear rates no longer collapse on a mastercurve upon horizontal rescaling. 

\end{appendices}

\bibliography{sn-bibliography}

\end{document}